\documentclass[11pt,a4paper]{article}
\usepackage{jheppub,multirow}
\usepackage{float}
\usepackage{graphicx}
\usepackage{url}
\usepackage{cancel}
\usepackage{slashed}
\usepackage{placeins}
\usepackage{feynman}
\usepackage{soul} 
\usepackage{bm}
\usepackage{amsmath}
\usepackage{amssymb}
\usepackage{listings}
\usepackage[toc, page]{appendix}
\usepackage{color}
\usepackage{booktabs} 
\usepackage{pgf}
\usepackage{empheq}
\usepackage{tikz-feynman}
\usepackage{tikz}

\usepackage{lipsum}
\usepackage{tabularx}
 \usetikzlibrary{decorations.pathmorphing}
  \usetikzlibrary{decorations.markings}
\tikzset{snake it/.style={decorate, decoration=snake}}
\usetikzlibrary{shapes.geometric, arrows}
\tikzstyle{process} = [rectangle, minimum width=3cm, minimum height=1cm, text centered, draw=black, fill=orange!30]
\tikzstyle{arrow} = [thick,->,>=stealth]
\usetikzlibrary{arrows.meta,bending}
\usepackage{subfigure}
\usepackage{pgf}
\usetikzlibrary{arrows}
\usetikzlibrary{patterns}
\renewcommand{\vec}[1]{\textbf{#1}}
 \usepackage{braket}
 \usepackage[inline]{enumitem} 
 
\usepackage{pifont}
\usepackage{array}
\newcolumntype{P}[1]{>{\centering\arraybackslash}p{#1}}
\usepackage{amsthm}
\usepackage{hyperref}
 \hypersetup{
    colorlinks=true,
    linkcolor=blue,
    filecolor=black,      
    urlcolor=blue,
}
\usepackage{tensor}
\usepackage{slashed} 

\def\G1{{\bf \gamma^{(1)}_N}}

\definecolor{violet}{cmyk}{0,1,0,0.2}

\usepackage[font=small,labelfont=bf,labelsep=period]{caption}
\usepackage{tabularx}
\newcolumntype{C}[1]{>{\centering\arraybackslash}p{#1}}

\newcommand{\be}{\begin{equation}}
\newcommand{\ee}{\end{equation}}
\newcommand{\bea}{\begin{eqnarray}}
\newcommand{\eea}{\end{eqnarray}}
\newcommand{\bi}{\begin{itemize}}
\newcommand{\ei}{\end{itemize}}
\newcommand{\ben}{\begin{enumerate}}
\newcommand{\een}{\end{enumerate}}

\numberwithin{equation}{section}
\numberwithin{figure}{section}
\numberwithin{table}{section}

\title{The dark side of the proton}
\author[a]{Matthew McCullough,}
\author[b]{James Moore,}
\author[b]{Maria Ubiali}
\affiliation[a]{CERN, Theoretical Physics Department, Geneva, Switzerland}
\affiliation[b]{DAMTP, University of Cambridge, Wilberforce Road,
  Cambridge, CB3 0WA, United Kingdom}
\emailAdd{matthew.mccullough@cern.ch}
\emailAdd{james.moore@damtp.cam.ac.uk}
\emailAdd{M.Ubiali@damtp.cam.ac.uk}

\abstract{We study the sensitivity of the High-Luminosity LHC to a light baryonic
  dark photon $B$, primarily coupled to quarks, as a constituent of
  the proton. This is achieved by allowing for a dark photon parton
  distribution function (PDF)
  in the PDF evolution equations. Depending on the mass and
  coupling of the dark photon, the evolution of standard quark and
  gluon PDFs is distorted to varying degrees.
  By analysing the effect of the dark photon on the tails of Drell-Yan
  invariant mass distributions, we demonstrate the potential of
  the LHC in determining competitive bounds on dark photon
  parameter space.
}

\arxivnumber{2203.xxxxx}
\subheader{\small CERN-TH-2022-043}

\begin{document}

\maketitle
\flushbottom

\section{Introduction}
The majority of the visible sector cosmological energy budget is comprised of hadrons,
yet it is rendered visible by the photon, which itself makes up only a tiny fraction of the energy
budget and does not behave as matter.  It is not unreasonable to expect that the moment the curtains
to the dark sector are drawn back it will be rays of dark light that flood detectors and not necessarily
the dominant matter component.  Thus, the most effective strategy to unveil the particle physics of
the dark sector may be to search for new light states carrying a vanishingly small fraction of the dark
energy budget;  perhaps, even, dark photons (hereafter referred to as `$B$').  In recent years searches
for dark photons have gained momentum, both theoretically and experimentally; see, for example, the recent reviews \cite{Battaglieri:2017aum,Fabbrichesi:2020wbt,Graham:2021ggy}, which paint a picture of the breadth of activities in this area.

Being Abelian vectors, dark photons can naturally be light, thus no specific mass scale is particularly
deserving of attention than another.  As a result experimental search strategies should endeavour to
cover as broad a mass range as possible.  Below $m_B \lesssim 1$ GeV a variety of intensity frontier
experiments have significant sensitivity to the presence of dark photons, however above this mass
scale only high energy accelerators have the capability to probe dark photon parameters.

Pursuing this program, \cite{Kribs:2020vyk,Thomas:2021lub} use deep
inelastic scattering (DIS) data from HERA, and projected data at the
Large Hadron Electron Collider (LHeC), to derive bounds on a particular class of dark photon models, in which the dark photon is introduced via kinetic mixing with the
SM electroweak bosons. In these studies, the dark photon is treated as a mediator of DIS,
hence modifying the theoretical expressions for the DIS structure functions, which allows
for the extraction of bounds.
Further, in~\cite{Thomas:2021lub} it is noted that a fully-consistent treatment using this approach requires a simultaneous fit of both parton 
distribution functions (PDFs) and dark photon parameters;
here, the interplay is a mild second-order effect, yielding a small relaxation of the constraints derived in~\cite{Kribs:2020vyk} (however, in~\cite{greljo_parton_2021,Iranipour:2022iak} it was shown that at the reach and precision of the high-luminosity phase of the Large Hadron Collider (HL-LHC), simultaneous 
analysis of PDFs and BSM effects will be significantly more impactful). 

What if a dark photon was baryonic, being primarily coupled to quarks over leptons? In this case, PDF effects take centre-stage, and 
it becomes reasonable to consider the dark photon not simply as a mediator of DIS, but as a \textit{constituent} of the proton in its own right. 

This is not without precedent; whilst the vast majority of New Physics (NP) searches at the LHC
involve processes initiated by coloured partons, namely
quarks and gluons, it is well-known that quantum fluctuations can give rise to non-coloured
partons inside hadrons, although with much smaller abundance. A key example is the inclusion 
of photons and leptons as constituents of the proton, which can play a
crucial role in achieving precise phenomenological
predictions at the LHC. In the recent {\tt LUXqed} publication it was shown that the photon PDF can
be determined in a model-independent manner, using DIS structure
function data~\cite{Manohar:2016nzj,Manohar:2017eqh}. These results brought an extremely accurate
determination of the photon PDF, that superseded the previous model-driven or purely data-driven analyses~\cite{Martin:2005,Ball:2013}; now, the {\tt LUXqed} method has been incorporated in several global PDF sets~\cite{Bertone:2017bme,Cridge:2021pxm,Guzzi:2021rvo}.
Going beyond just photon PDFs, the {\tt LUXqed} approach has since been extended to the computation
of $W$ and $Z$ boson PDFs~\cite{Fornal:2018znf}, and lepton PDFs~\cite{Buonocore:2020nai}.
Whilst the impact of the photon PDFs is sizeable in a number of
kinematic regions, the impact of lepton PDFs is rather small at Run III. However
lepton-initiated processes will become an important feature in the near future, 
in particular in the HL-LHC phase, which  
will provide the largest proportion of new high-energy particle physics data in the next 20 years~\cite{Buonocore:2020erb,Greljo:2020tgv,Buonocore:2021bsf,Harland-Lang:2021zvr}.

In this spirit, we might reasonably ask whether the proton could contain small contributions from a dark photon, the consideration of which
could be important in the near future. In this work, we assess the impact of the inclusion in the proton
of a new, light baryonic dark photon $B$ with mass in the 
range $m_B \in [2,80]\ \text{GeV}$, coupling primarily to quarks via
the effective interaction Lagrangian:
\begin{equation}
\label{eq:darkphotoninteraction}
\mathcal{L}_{\text{int}} = \frac{1}{3}g_B \bar{q} \slashed{B} q,
\end{equation}
where the dark fine structure constant is of the order $\alpha_B \sim 10^{-3}$.
The dark photon's parton distribution enters into the PDF evolution equations 
in the same way as the photon PDF, except for a flavour-universal
coupling and a non-zero mass threshold. 
The other PDFs, particularly the quarks and antiquarks, are modified by the presence of a dark photon,
especially in the large-$x$ region; this gives rise to significantly
different predictions for key observables that can be measured at a
very high degree of precision at the LHC. In this work we focus on the
high invariant-mass Drell-Yan (DY) differential distributions, whose
theoretical description is significantly affected by the distortion of
the quark and antiquark PDFs due to the presence of a non-zero dark
partonic density.

For the first time in the literature, we demonstrate the strong
sensitivity to this dark photon's mass and coupling
of the precise measurements of the high-mass Drell-Yan tails at the HL-LHC, by looking at the
data-theory agreement using the standard PDFs and the PDFs modified by the presence of a non-zero
dark photon distribution.
Whilst the sensitivity of collider measurements to BSM colored
partons in the proton has been shown to be very
strong~\cite{Berger:2004mj,Berger:2010rj} -- as one would expect given that light coloured
particles very rapidly distort both the DGLAP evolution and the
running of $\alpha_s(\mu)$~\cite{Becciolini:2014lya} --
in this work, we show that in the near future we will still be able to competitively probe the presence of a dark parton that couples to
quarks via a much more subtle QED-like mixing.

In Sec.~\ref{sec:dglap} we describe how PDF evolution is modified by the presence of a non-zero
dark photon distribution, and state the order of our calculation in
QCD and QED perturbation theory. Additionally, we show the dark photon
distributions that we obtain, and display how the other parton distributions
are modified by the presence of the dark photon. 
Following this, in Sec.~\ref{sec:pheno} we consider how the presence
of the dark photon PDF affects precision theoretical predictions for DY processes at the HL-LHC.
If observations at the HL-LHC are SM-like, this could be used to place bounds on the dark photon content of the proton
and hence constrain the parameter space of this model.  In certain mass ranges we find that these projected bounds
are competitive with existing limits, demonstrating the extraordinary ability of the precision-QCD era of LHC physics
to probe new light dark sector physics.

\section{Determination of the dark PDF contribution}
\label{sec:dglap}
In order to produce a PDF set which include a non-zero dark photon distribution,
we follow the method described in \cite{Bertone:2015lqa},
which constitutes a first exploration into the effects of the inclusion of lepton PDFs.
In this study, 
simple ans\"{a}tze for the functional forms of the light lepton PDFs (electrons and muons)
are postulated at the initial PDF parametrisation scale $Q_0 = 1.65\ \text{GeV}$, based on the assumption that
initial-state leptons are primarily generated by photon splitting, 
while leptons that are heavier than the initial-scale (namely the
tauon) are dynamically generated at their mass threshold and
kinematical mass effects are neglected, as is done for all heavy
partons in the Zero-Mass Variable-Flavour-Number scheme (ZMVFN)~\cite{Maltoni:2012pa,Bertone:2017djs}.  
All parton flavours, including the lepton ans\"{a}tze alongside initial quark,
gluon and photon PDFs drawn from some fixed baseline PDF fit, are then
evolved using an appropriately modified version of the PDF evolution equations,
the \textit{Dokshitzer-Gribov-Lipatov-Altarelli-Parisi} (DGLAP)
equations~\cite{Altarelli:1977zs,Gribov:1972ri,Dokshitzer:1977sg}, thus producing
a final PDF set now including lepton PDFs. 

Analogously in this work, given that the dark photon mass range that
we consider here is above 2 GeV,
we dynamically generate a PDF at the dark photon mass threshold $m_B$. 
We then evolve this new distribution alongside quark, antiquark,
gluon and photon PDFs drawn from a baseline set.
Hence, via the interplay between the flavours generated by
the DGLAP evolution, the resulting quark, gluon and photon PDFs differ relative
to the original reference set evolved excluding dark photons, allowing the impact
of the dark photon inclusion to be assessed (and, in the subsequent section, bounds from HL-LHC
projected pseudodata to be extracted).

In this section, we describe this procedure in more detail.
We begin by explicitly showing the modification to the DGLAP equations
required by the presence of dark photons in the proton.
We then display and discuss the resulting `dark PDF sets',
and compare them to baseline PDF sets excluding the dark photon.
In particular, we analyse the dark luminosities, which show an appreciable deviation
from their standard counterparts for sufficiently large values of the coupling $\alpha_B$;
this motivates the phenomenological study that we present in Sec.~\ref{sec:pheno}.

\subsection{The DGLAP equations in the presence of dark photons}
\label{subsec:dglap}


As is well known, in order to combine  QCD and electroweak
calculations at hadron colliders,
the PDF evolution must be determined using the coupled QCD$\otimes$QED
DGLAP
evolution
equations~\cite{DeRujula:1979grv,Kripfganz:1988bd,Blumlein:1989gk}.
Here, we modify these equations by adding the leading order evolution of a dark photon PDF. 
In order to assess the impact of such a dark photon PDF in the
evolution, it is essential to include all QCD and QED contributions
of the same magnitude as the leading dark contribution. Indeed, to
include the terms multiplied by $\alpha_B \sim 10^{-3}$ consistently,
we must also include the terms multiplied by $\alpha_s \sim 10^{-1}$,
$\alpha_s^2 \sim 10^{-2}$, $\alpha_s^3 \sim 10^{-3}$, $\alpha \sim
10^{-2}$ and $\alpha\alpha_s \sim 10^{-3}$
in the evolution (and further it is no loss to include the terms
multiplied by $\alpha^2 \sim 10^{-4}$);
in particular, we work at NNLO in QCD, NLO in QED, and include QCD-QED
interference;
furthermore, we always include a photon PDF. On the other hand, the
lepton PDFs determined
in~\cite{Buonocore:2020erb,Buonocore:2020nai,Buonocore:2021bsf,Harland-Lang:2021zvr}
give a contribution that is more than one order of magnitude smaller
than the dark photon
contributions determined in this work, thus we can safely ignore them.

With the orders and flavours specified, the modified DGLAP equations which we use in this work can be stated as: 
\begin{align}
\label{eq:basicdglap}
\mu^2 \frac{\partial g}{\partial \mu^2} &= \sum_{j=1}^{n_f} P_{gq_j} \otimes q_j +\sum_{j=1}^{n_f} P_{g\bar{q}_j} \otimes \bar{q}_j + P_{gg}\otimes g + P_{g\gamma}\otimes \gamma \notag\\[1.5ex]
\mu^2 \frac{\partial \gamma}{\partial \mu^2} &= \sum_{j=1}^{n_f} P_{\gamma q_j} \otimes q_j +\sum_{j=1}^{n_f} P_{\gamma\bar{q}_j} \otimes \bar{q}_j + P_{\gamma g}\otimes g + P_{\gamma\gamma}\otimes\gamma\notag \\
\\[-0.5ex]
\mu^2 \frac{\partial q_i}{\partial \mu^2} &= \sum_{j=1}^{n_f} P_{q_iq_j} \otimes q_j +\sum_{j=1}^{n_f} P_{q_i\bar{q}_j} \otimes \bar{q}_j + P_{q_ig}\otimes g + P_{q_i\gamma}\otimes \gamma + P_{q_iB} \otimes B \notag\\[1.5ex]
\mu^2 \frac{\partial B}{\partial \mu^2} &= \sum_{j=1}^{n_f} P_{Bq_j} \otimes q_j +\sum_{j=1}^{n_f} P_{B\bar{q}_j} \otimes \bar{q}_j + P_{BB}\otimes B,\notag
\end{align}
where $\mu^2$ is the factorisation scale, $n_f$ the number of active
flavours, $q_i$ ($\overline{q}_i$) the parton density of the $i^{\rm
  th}$ (anti)quark, $g$ the gluon PDF, $\gamma$ the photon PDF, and
$B$ the new dark photon PDF. The symbol $\otimes$ denotes the usual Mellin convolution:
\begin{equation}
\label{eq:convolution}
(f \otimes g)(x) = \int\limits_{x}^{1} \frac{dy}{y} f(y) g\left( \frac{x}{y} \right).
\end{equation}
 Evolution equations for antiquarks can be obtained by employing conjugation
invariance. 

The matrix elements $P_{ij}$ (with $i,j=q,\bar{q},g,\gamma,B$) are perturbatively calculable functions\footnote{Technically, mathematical distributions.} called \textit{splitting functions}. We can decompose the splitting functions into series of the form:
\begin{align}
\label{eq:splittingexpansion}
P_{ij} &= \left( \frac{\alpha_s}{2\pi} \right) P_{ij}^{(1,0,0)} + \left( \frac{\alpha_s}{2\pi} \right)^2 P_{ij}^{(2,0,0)} + \left(\frac{\alpha_s}{2\pi}\right)^3 P_{ij}^{(3,0,0)} \notag\\[1.5ex]
&\quad+ \left( \frac{\alpha}{2\pi} \right) P_{ij}^{(0,1,0)} + \left( \frac{\alpha_s}{2\pi} \right) \left( \frac{\alpha}{2\pi} \right) P_{ij}^{(1,1,0)} + \left( \frac{\alpha}{2\pi} \right)^2 P_{ij}^{(0,2,0)} \\[1.5ex]
&\quad+ \left( \frac{\alpha_B}{2\pi} \right) P_{ij}^{(0,0,1)} + \cdots, \notag
\end{align}
where we follow the notation of \cite{deFlorian:2015ujt,deFlorian:2016gvk}; the upper indices indicate
the (QCD,QED,Dark) order of the calculation (where in this work we have added an additional `Dark' index, corresponding
to the powers of the dark coupling $\alpha_B$). The QCD contributions
to the splitting functions were fully computed up to $O(\alpha_s^3)$
in~\cite{Vogt:2004mw,Moch:2004pa}\footnote{They are also partially
  known at $O(\alpha_s^4)$~\cite{Moch:2021qrk}, but this contribution
  are not yet fully known and are not included in any public 
PDF evolution code. Furthermore,
their contribution is beyond the accuracy needed in the current analysis.}, the mixed QED and QCD contribution $P_{ij}^{(1,1,0)}$
was computed in~\cite{deFlorian:2015ujt}, and the NLO QED contribution $P_{ij}^{(0,2,0)}$ was computed in~\cite{deFlorian:2016gvk}.

\begin{figure}[t]
\centering
 \subfigure[]{
\begin{tikzpicture}[scale=0.8]
\begin{scope}[thick,decoration={
    markings,
    mark=at position 0.5 with {\arrow{>}}}
    ] 
\draw[postaction = {decorate}] (-1,-1) node[below left]{$q$} -- (0,0);
\draw[postaction = {decorate}] (0,0) -- (1,1) node[above right]{$q$};
\end{scope}

\draw[thick, snake it] (0,0) -- (1.5,0) node[right]{$B$};
\end{tikzpicture}}
\qquad
 \subfigure[]{
\begin{tikzpicture}[scale=0.8]
\begin{scope}[thick,decoration={
    markings,
    mark=at position 0.5 with {\arrow{>}}}
    ] 
\draw[postaction = {decorate}] (-1,-1) node[below left]{$q$} -- (0,0);
\draw[postaction = {decorate}] (0,0) -- (1.5,0) node[right]{$q$};
\end{scope}

\draw[thick, snake it] (0,0) -- (1,1) node[above right]{$B$};
\end{tikzpicture}}
\qquad
 \subfigure[]{
\begin{tikzpicture}[scale=0.8]
\begin{scope}[thick,decoration={
    markings,
    mark=at position 0.5 with {\arrow{>}}}
    ] 
\draw[postaction = {decorate}] (1,1)  node[above right]{$\bar{q}$}-- (0,0) ;
\draw[postaction = {decorate}] (0,0)  -- (1.5,0) node[right]{$q$};
\end{scope}

\draw[thick, snake it] (-1,-1) node[below left]{$B$} -- (0,0) ;
\end{tikzpicture}}
\qquad
 \subfigure[]{
\begin{tikzpicture}[scale=0.8]
\draw[thick, snake it] (-1,-1) node[below left]{$B$} -- (1,1) node[above right]{$B$};
\end{tikzpicture}}
\caption{The diagrams involving dark photons which contribute to splitting functions. (a), (b), (c), (d) show the contributions to $P_{qq}^{(0,0,1)}(x)$, $P_{Bq}^{(0,0,1)}(x)$, $P_{qB}^{(0,0,1)}(x)$ and $P_{BB}^{(0,0,1)}(x)$, respectively (note at this order, $P_{BB}^{(0,0,1)}(x)$ is proportional to a delta function, $\delta(1-x)$, indicating the lack of possible splitting in this channel).}
\label{fig:splittingfunctions}
\end{figure}
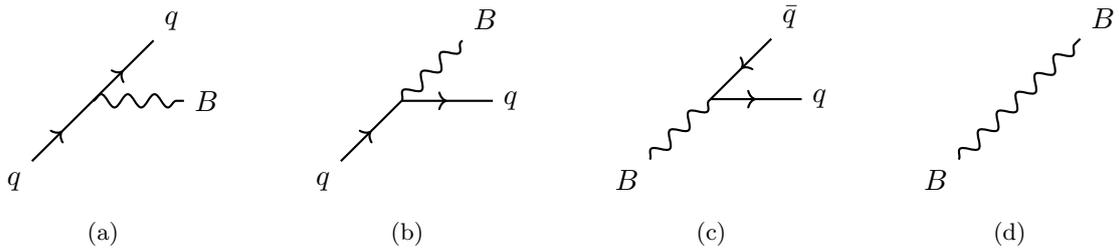

The coefficients $P_{ij}^{(0,0,1)}$ can be calculated directly by
finding the most collinearly-divergent parts of the four dark
splitting channels pictured in Fig.~\ref{fig:splittingfunctions}. The only non-zero contributions are given by $ij=qq, qB, Bq$ and $BB$ (the same results for the antiquarks can be obtained by charge conjugation).
We briefly summarise the calculation in Appendix~\ref{app:splitting}.
However, a detailed calculation is not strictly necessary, since the form
of the interaction Lagrangian Eq.~(\ref{eq:darkphotoninteraction}) is identical
to that of the electromagnetic-hadronic interaction in the SM,
except with a universal coupling $\frac{1}{3}g_B$ to all quarks and antiquarks.
It follows that the splitting function contributions provided by the
dark photon $B$ will be identical (up to a factor of $\frac{1}{9}$, due to
our convention for the universal coupling) to those provided by the
photon $\gamma$; in particular, we can quote the required leading-order
splitting functions by comparing to \cite{Bertone:2015lqa}:
\begin{gather}
\label{eq:splittingfunctions}
P_{qq}^{(0,0,1)}(x) = \frac{1+x^2}{9(1-x)_+} + \frac{1}{6} \delta(1-x), \qquad P_{BB}^{(0,0,1)}(x) = -\frac{2}{27}\delta(1-x), \notag\\ \\
P_{qB}^{(0,0,1)}(x) = \frac{x^2 + (1-x)^2}{9},\qquad P_{Bq}^{(0,0,1)}(x) = \frac{1}{9} \left( \frac{1 + (1-x)^2}{x} \right). \notag
\end{gather}
Here, the $+$ notation used in $P_{qq}^{(0,0,1)}(x)$ denotes
the usual \textit{plus-distribution}, defined by:
\begin{equation}
\label{eq:plusdistribution}
\int\limits_{0}^{1} f(x)_+ g(x)\ dx := \int \limits_{0}^{1} f(x) (g(x) - g(1)).
\end{equation}

To solve the modified DGLAP equations~\eqref{eq:basicdglap}, we must
also specify initial conditions for the dark photon at the initial scale $Q_0 = 1.65\ \text{GeV}$. 
Given that we consider masses $m_B \in [2,80]$ GeV, we set $B(x,Q^2)
= 0$ for all $Q < m_B$, and we generate the dark photon PDF dynamically
at the threshold $Q = m_B$ from PDF evolution, similar to the
treatment of heavy quarks in the ZM-VFN scheme~\cite{Maltoni:2012pa,Bertone:2017djs}, and the tauon
PDF in~\cite{Bertone:2015lqa}. Hence the dark photon PDF is always
proportional to the dark photon coupling $\alpha_B$ and to
$\log(Q^2/m_B^2)$ for $Q>m_B$. 

\subsection{PDF sets with dark photons}

We have implemented the modified DGLAP equations described in~Sec. \ref{subsec:dglap} in the
public \texttt{APFEL} PDF evolution code; more detail regarding the code 
implementation is given in App.~\ref{app:code}. Using
the modified code, we produce a PDF set and a corresponding LHAPDF
grid~\cite{Buckley:2014ana} including dark photons, for each given
value of the dark photon mass and coupling that we consider. 
We focus on the introduction of a dark photon into the evolution of the
{\tt NNPDF3.1luxQED} set~\cite{Bertone:2017bme}\footnote{This
    set will be soon superseded by the PDF set including QED effects
    obtained starting from NNPDF4.0~\cite{Ball:2021leu}.}, which provides our
SM baseline, namely an NNLO
global PDF analysis of all standard parton flavours together with a photon PDF (the photon
PDF in this set is determined using the {\tt LUXqed}
method~\cite{Manohar:2017eqh}).

In this section, we display the key
results from a `dark PDF set' in a particular scenario
that is permitted according to the bounds given in
Ref.~\cite{Ilten:2018crw}, namely:
\begin{equation}
  \label{eq:value}
  m_B = 50\,\, \text{GeV}, \qquad \alpha_B = 3 \times 10^{-3},
\end{equation}
which corresponds to taking $g_B = 1.94 \times 10^{-1}$. As described
above, a massless dark photon is generated dynamically at the threshold $Q = m_B$, and is set
to zero before this threshold is reached. We have chosen a sufficiently
high (admissible) value of the coupling to display the impact
upon PDFs and parton luminosities. 
\begin{figure}[hbt]
\centering
\includegraphics[width=0.49\textwidth]{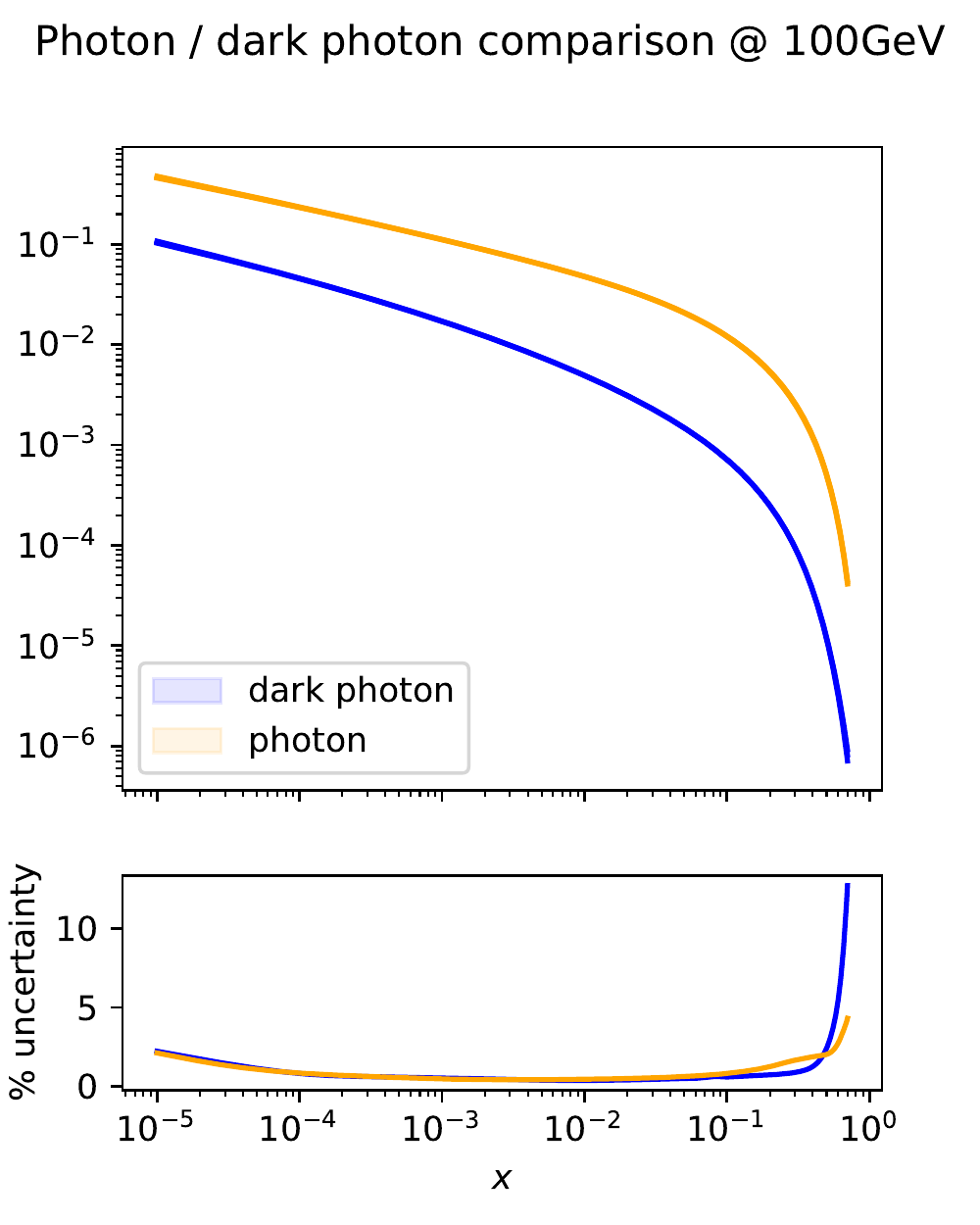}
\includegraphics[width=0.49\textwidth]{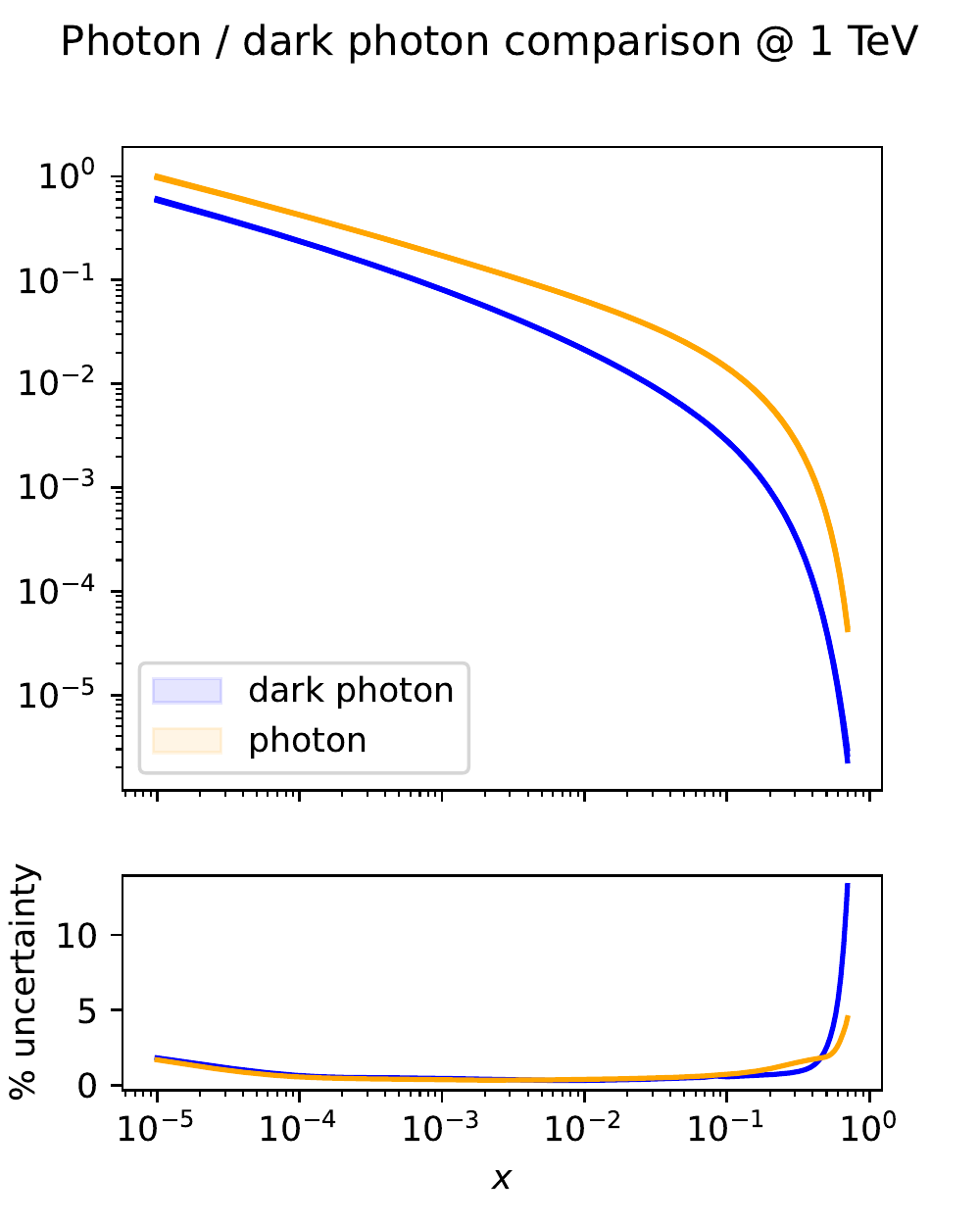}
\caption{Comparison of $x\gamma(x,Q^2)$ and $xB(x,Q^2)$ at $Q=100\ \text{GeV}$ (left) and $Q=1\ \text{TeV}$ (right)
for the values of dark photon mass and coupling given in
Eq.~\eqref{eq:value}. The percentage relative 68\% C.L. PDF
uncertainties of the photon and the dark photon are
  displayed in the bottom inset.}
\label{fig:dark_pdfs}
\end{figure}

In Fig.~\ref{fig:dark_pdfs}, we display both the photon and dark
photon PDFs in our representative dark set (obtained by setting the
dark photon coupling and mass at the values given in
Eq.~\eqref{eq:value}) at the scales $Q = 100\ \text{GeV}$ and $Q=1\ \text{TeV}$,
and show their relative PDF uncertainties. As anticipated, the dark photon PDF features 
the same functional form as the photon PDF (this is to be expected
since the photon and dark photon splitting functions are identical up
to scaling), but its density is smaller since $\alpha_B \lesssim
\alpha$. Furthermore, it can be shown that increasing $\alpha_B$, and also moderately decreasing 
$m_B$, increases the similarity of the dark photon and photon PDFs.
The dark photon uncertainty is mostly comparable to the photon
uncertainty up to $x\sim 0.4$, and then increases faster than the photon
uncertainty. This is due to the dark photon being generated off the
singlet PDF (the sum of all quarks and antiquarks) at its mass
threshold with a rather small coupling; in particular, the dark
photon uncertainty is comparable to the uncertainty of the singlet PDF
scaled by a factor of  $\alpha_B$. This makes it comparable to the
photon PDF uncertainty (for the choice of $\alpha_B$ and $m_B$ of
Eq.~\eqref{eq:value}), except in the large-$x$ region where the singlet PDF uncertainty dramatically increases,
resulting in the dark photon PDF uncertainty to consistently increase up to $\sim10\%$ at
$x\sim 0.6$. We have verified that for larger couplings the
uncertainty increases, as one would expect.

Now that we have introduced a new parton in the proton, it is 
interesting to ask how much `space' it takes up; this can be quantified
by determining the momentum carried by the dark photon at 
different energy scales. As usual, the momentum fraction carried by any given
parton flavour $f$ at the scale $Q$ is defined by:
\begin{equation}
\label{eq:momfraction}
\langle x\rangle_f(Q) := \int\limits_{0}^{1} dx\ xf(x,Q).
\end{equation}
In Table.~\ref{tab:dark_momentum}, we give a comparison between the
momentum carried by the dark photon, the photon and the singlet for
the representative dark PDF
set computed using the values specified in Eq.~\eqref{eq:value}, and
compare them to the baseline SM PDF set, at $Q=100$ GeV and $Q=1$ TeV. 
We observe that the fraction of the proton momentum carried by the
dark photon increases with the scale $Q$, which is to be expected 
by analogy with the photon's behaviour. Depending on the coupling
and the mass of the dark photon, the latter carries up to a fraction
of percent of the proton momentum's fraction at $Q\approx1\
\text{TeV}$. 
\begin{table}[tb]
\centering
\begin{tabularx}{\textwidth}{X|C{1.8cm}C{1.8cm}C{1.8cm}C{1.8cm}}
\toprule
$\langle x\rangle_f(Q=100$ GeV) & $f=\Sigma$ & $f=\gamma$ & $f=B$ \\
\midrule
Baseline &  50.23\% &  0.4241\% & 0\% \\
Dark set & 50.17\% & 0.4241\% &0.03214\%  \\
\midrule
$\langle x\rangle_f(Q=1$ TeV) & $f=\Sigma$ & $f=\gamma$ & $f=B$ \\
\midrule
Baseline &  48.36\% &  0.5279\% & 0\% \\
Dark set & 48.12\% &  0.5275\% & 0.1357\%  \\
\bottomrule
\end{tabularx}
\caption{A comparison between the momentum fraction percentage carried by the
  singlet $\Sigma$, the photon $\gamma$, and the dark photon $B$ at $Q=100$ GeV and $Q=1$ TeV,
  for the baseline SM set and the dark
  PDF set, obtained with the photon coupling and mass given in Eq.~\eqref{eq:value}. The momentum
fraction is computed on the central replica in each case. }
\label{tab:dark_momentum}
\end{table}

Crucially, the presence of a dark photon in the DGLAP equations also
modifies the evolution of all other flavours of PDFs due to the
coupling of the PDFs via the modified DGLAP equations
Eq.~(\ref{eq:basicdglap}).
We expect that the modification of the quark and antiquark 
flavours is strongest, as the dark photon is directly coupled to
them. We also anticipate a modification to the gluon and photon
PDFs, but these will be second order effects, so we expect that they
will be smaller in comparison. Moreover, the density of each of the
flavours should reduce, as the new dark photon `takes up space' in the
proton which was previously occupied by the other flavours.
\begin{figure}[tb]
\centering
\includegraphics[width=0.49\textwidth]{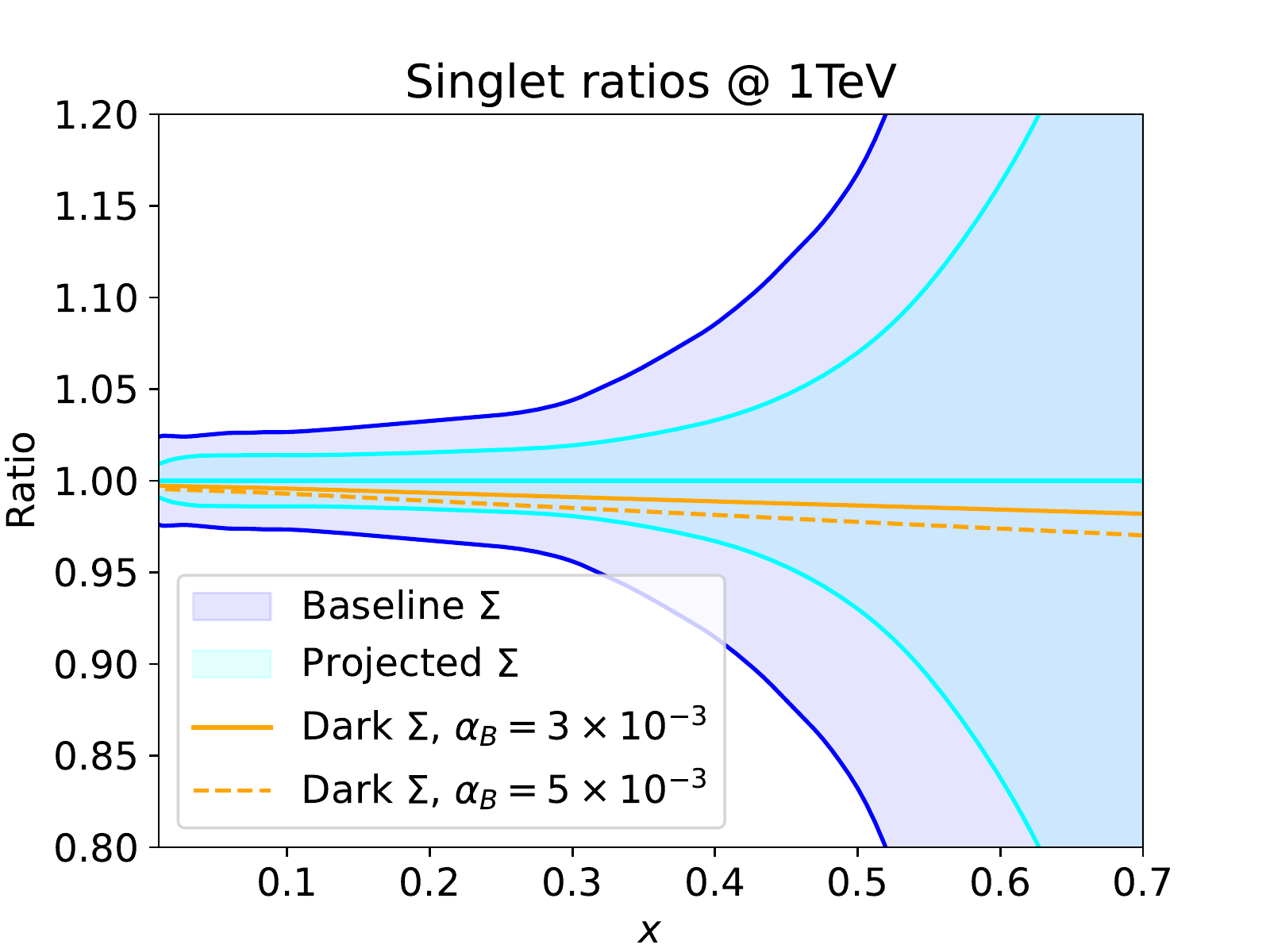}
\includegraphics[width=0.49\textwidth]{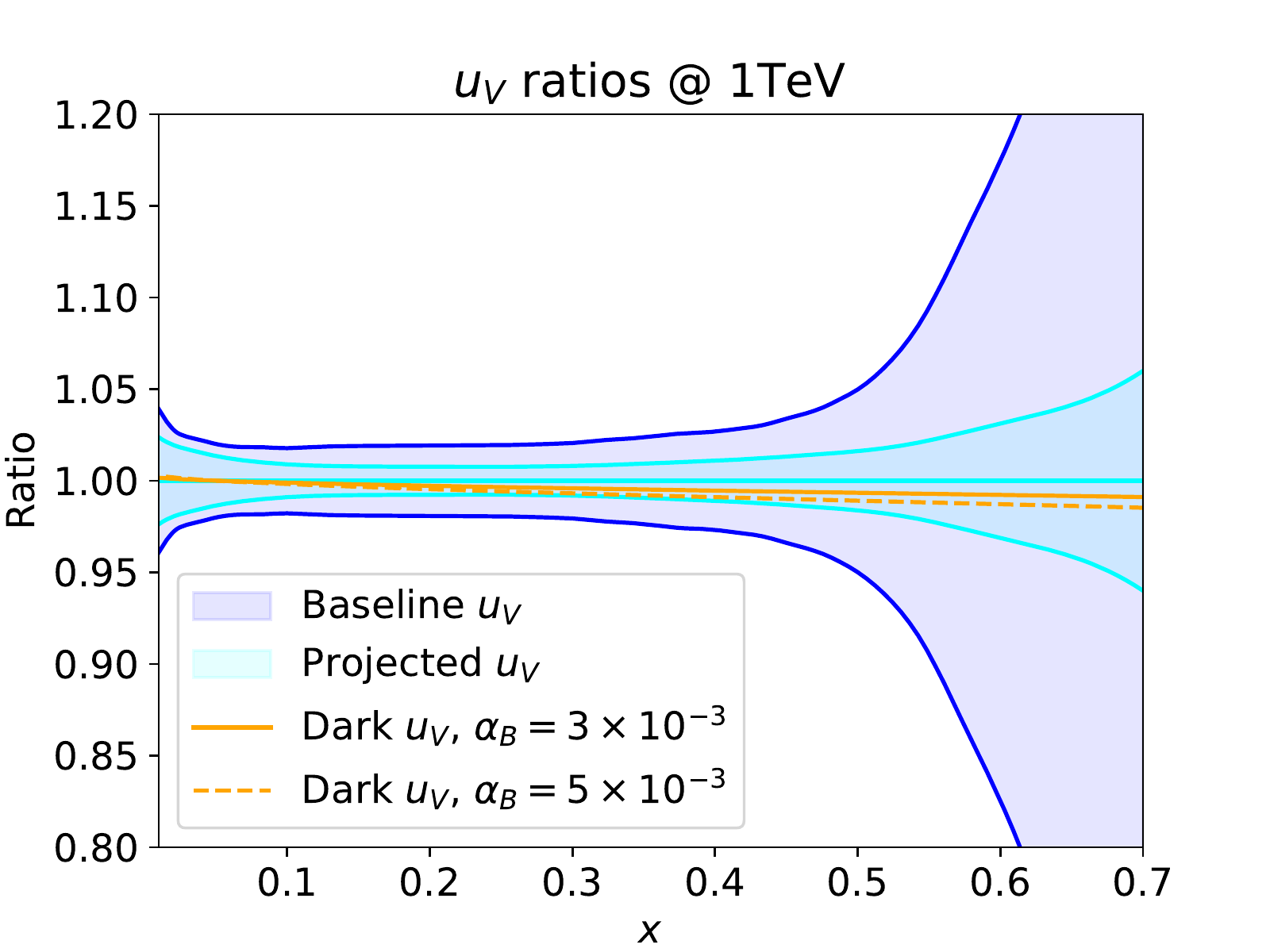}
\caption{In solid orange, the ratio between the central singlet PDF $\Sigma$ (left) and central $u$-valence
  PDF (right), drawn from the dark benchmark scenario in Eq.~\eqref{eq:value}, 
  to the baseline SM central replica at $Q = 1\
  \text{TeV}$. In dashed orange, the same ratios but between the SM baseline and a dark PDF set
  produced using $m_B = 50\ \text{GeV}, \alpha_B = 5 \times 10^{-3}$. In each case, the uncertainty bands
  represent the 68\% C.L. PDF uncertainties of the baseline set (in blue) 
  and the projected PDF uncertainties at the HL-LHC, determined
  from Ref.~\cite{Khalek:2018} (in light blue). The deviation when $\alpha_B = 5 \times 10^{-3}$ approaches
  the boundary of the projected HL-LHC uncertainty bands, consistent with the behaviour we see in Fig.~\ref{fig:14tevlumis} later; increasing $\alpha_B$ (and also to a milder extent, decreasing $m_B$) pushes the deviation outside of projected HL-LHC uncertainty bands.
  See the main text for more details.}
\label{fig:ratio_quarks}
\end{figure}
Results are shown in Fig.~\ref{fig:ratio_quarks}, in which the ratio
between the central value of the dark-photon modified singlet
($u$-valence) PDF and the central value of the baseline singlet
($u$-valence) PDF are displayed and compared to the current 68\% C.L. PDF
uncertainty. 

We observe
that the modification of the singlet becomes visible at about $x\sim 0.2$ and
reaches 3\% at larger values of $x\sim 0.5$. This is well within the
68\% C.L. uncertainty of the singlet PDF from the baseline {\tt
  NNPDF3.1luxQED} NNLO set. However, thanks to the inclusion of a vast
number of new datasets and the increased precision of the methodology
used in global PDF analysis, the recent {\tt NNPDF4.0} NNLO
set~\cite{Ball:2021leu} 
displays significantly smaller large-$x$ uncertainty. Such a
decrease in PDF uncertainties goes in the direction indicated by the
dedicated study on how PDF
uncertainties will decrease in future, thanks to the inclusion of precise
HL-LHC measurements~\cite{Khalek:2018}. In particular, to give an indication
of how the modification of PDFs due to the presence of a dark photon might
come into tension with decreasing PDF uncertainties during the HL-LHC phase, 
we display the projected 68\% PDF uncertainties at
the HL-LHC determined in the `optimistic' scenario, Scenario 3, of Ref.~\cite{Khalek:2018}.
In this case, should PDF uncertainties decrease to the
level predicted by Ref.~\cite{Khalek:2018}, the distorted singlet PDF approaches the edge of the
projected PDF uncertainty at $x\sim \,0.1-0.3$ region, for the values given
in Eq.~\eqref{eq:value}. This is particularly relevant for the
analysis that we present in the next section. 

%


\section{Probing the Dark Sector}
\label{sec:pheno}

In this section we review the existing constraints on the dark photon.
Subsequently, in order to assess the impact of a non-zero dark photon
parton density on physical observables, we plot the parton luminosities
when the dark photon is included, as compared to our baseline SM set. 
We compare the predicted deviations with the current PDF
uncertainties and with the projected PDF uncertainties at the HL-LHC.
Finally, we motivate and present our analysis of projected HL-LHC Drell-Yan data 
and compare the maximal sensitivity we can achieve to the existing
bounds derived in the literature.

\subsection{Review of existing constraints on the dark photon}
\label{sec:existingconstraints}
To appreciate the utility of the dark photon PDF at colliders, we may compare to alternative probes.  Recent works considering this class of baryonic dark photon models include \cite{Dobrescu:2014fca,Dror:2017ehi,Ismail:2017fgq,Dror:2018wfl,Ilten:2018crw,Dobrescu:2021vak}.  There are a variety of competing constraints on this scenario, of varying theoretical robustness.

One class of constraints, first considered in detail in
\cite{Dobrescu:2014fca}, is theoretical and concerns the mixed $\text{U}(1)_B-$EW anomalies.  Suppose we envisage that the UV-completion of the model Eq.~\eqref{eq:value} is perturbative with $\text{U}(1)_B$ linearly realised.  In that case the mixed anomaly must be UV-completed by some fermions with electroweak charges.  In this perturbative UV-completion they will obtain their mass from spontaneous $\text{U}(1)_B$-breaking.  As a result, they will be coupled to the longitudinal mode of $B$ and an additional Higgs-like scalar with a Yukawa coupling $\lambda \propto M_F/v_B$, where $M_F$ is the fermion mass, $v_B$ is the $\text{U}(1)_B$-breaking expectation value, and we have assumed three sets of fermions with the same charge ($1/3$) as the left-handed fermions, for simplicity.  On the other hand we have $g_B  \propto M_B/v_B$ following from the charge and symmetry-breaking vacuum expectation value.  As a result, we expect:
\begin{equation}
g_B \approx \frac{2 \lambda}{3} \frac{M_B}{M_F} ~~,
\end{equation}
where the precise numerical factors are taken from \cite{Dobrescu:2014fca}.  Thus, requiring perturbativity $\lambda \lesssim 4 \pi$ implies an upper bound on $g_B$, where the factor $1/3$ follows from the fact that each family of fermions is in triplicate to mirror the QCD multiplicity of the SM quarks.  This limit is shown as a dashed line in Fig.~\ref{fig:limits} where we have taken $M_F\geq 90$ GeV for the electroweak-charged fermions. 

However, a number of implicit assumptions have been made which can weaken upon further inspection.  To see this, consider cancelling the anomaly with $N$ copies of the above class of fermions.  In this case the limit becomes:
\begin{equation}
g_B \lesssim \frac{8 \pi}{3} \frac{N}{3} \frac{M_B}{M_F} ~~.
\end{equation}
Hence we see that this theoretical limit makes not only the assumption of a weakly-coupled UV-completion, but also depends on assumptions of minimality of the UV completion as well.  As a result, while this limit does guide the eye as to the nature of the UV-completion, it cannot be considered a strong theoretical limit on the model parameters.

The only truly model-independent theoretical limit comes from considering the scale at which the validity of the IR theory itself breaks down.  Given that the quark interactions are vector-like there is no possibility of tree-level unitarity violation in quark scattering mediated by $B$, thus we must look to quantum effects.  In this case the mixed-anomaly becomes relevant and renders the theory non-unitary unless \cite{Preskill:1990fr}:
\begin{equation}
g_B \lesssim \frac{(4 \pi)^2}{3 \alpha_W} \frac{M_B}{M_\Lambda} ~~,
\end{equation}
where $\alpha_W$ is the $\text{SU}(2)$ fine structure constant at the electroweak scale and $M_\Lambda $ is the energy scale at which the theory becomes strongly coupled.  Numerically this is
\begin{equation}
g_B \lesssim \frac{3M_B}{5 \text{ GeV}} \frac{10 \text{ TeV}}{M_\Lambda} ~~,
\end{equation}
which is too weak to be relevant for our purposes.  As a result we conclude that the effective theory considered here is valid throughout the energy scales under investigation. However, we note that, as shown in
\cite{Dobrescu:2021vak}, the mixing with the $Z$-boson is sensitive to
the details of the UV-completion; for this reason we restrict the mass range under
investigation to $m_B \leq 80\ \text{GeV}$, above which these UV-dependent effects can be important.  

There are three relevant classes of experimental constraints.  The first
concerns the exotic $Z$-boson decays $Z\to B\gamma$.  These constraints
were calculated in \cite{Dror:2017ehi} based on the LEP analysis for
$Z\to H\gamma$, $H\to$hadrons~\cite{L3:1996dmt}.\footnote{Note that
  this reference does not appear in~\cite{Dror:2017ehi}, but instead
  in~\cite{L3:1991kow,L3:1992kcg}, however the authors of
  \cite{Dror:2017ehi} have confirmed that the limits follow from a
  recasting of \cite{L3:1996dmt}.} This limit, relevant to the higher
mass range, is shown in red in Fig.~\ref{fig:limits}.  The second
class of constraints at lower masses concerns exotic $\Upsilon$ decays
\cite{Carone:1994aa,Aranda:1998fr}, where the constraint is dominated
by limits on $\Upsilon(1S)\to 2$ jets \cite{ARGUS:1986nzm}, shown in
blue in Fig.~\ref{fig:limits}. Finally, there are additional searches for
hadronically decaying resonances at hadron colliders
\cite{Dobrescu:2013cmh,Shimmin:2016vlc,ATLAS:2019itm,Dobrescu:2021vak}.
The strongest are from CMS $B+$ISR searches
\cite{CMS:2019xai,CMS:2019emo}, shown in yellow in
Fig.~\ref{fig:limits}. 

\subsection{Effects of the dark photon on parton luminosities}
In Sec.~\ref{sec:dglap}, we showed that the presence of a dark photon modifies
all other flavours of PDFs via the mixing associated with the
DGLAP evolution equations, with a modification that is proportional to
$\alpha_B$ and the logarithm of $m_B$. 
In order to assess the impact of a dark photon
parton density on physical observables, and thus extract the 
sensitivity that the LHC can achieve on the
parameters of the model, in the following subsection
we compare the size of the dark parton luminosities to luminosities
involving the other partons, and assess the impact of the dark photon
on the dominant partonic channels.

Parton luminosities are doubly differential quantities defined as:
\begin{equation}
\frac{d{\cal L}_{ij}}{dyd\tau}=f_i(x_1,Q)f_j(x_2,Q)\qquad x_{1,2}=\sqrt{\tau}\exp(\pm
y)\qquad \tau=\frac{M_X^2}{S},
\end{equation}
where $S$ is the squared centre-of-mass energy of the hadronic collision, $M_X$ is the invariant mass
of the partonic final state, $y$ is the rapidity of the partonic final state, and 
$f_i(x,Q)$ is the PDF of the $i^{\rm th}$ parton evaluated at the
scale $Q$. Different choices for $Q$ can be adopted in order to improve predictions
of a particular process and/or distribution.
At the level of pure luminosities, without the convolution with
any specific matrix element, the factorisation scale can be naturally
set to $Q = M_X$. For plotting purposes, it is useful to define the
$M_X$-differential luminosities, given by:
\begin{equation}
  \label{eq:lumidef}
\Phi_{ij}(M_X) \,=\,\frac{d{\cal L}_{ij}}{dM_X^2}=\frac{1}{S}\,\int\limits_{M_X^2/S}^1
\frac{dx}{x}\ f_i(x,M_X)f_j\left(\frac{M_X^2}{xS},M_X\right).
\end{equation}

\begin{figure}[tb]
\centering
\includegraphics[width=0.7\textwidth]{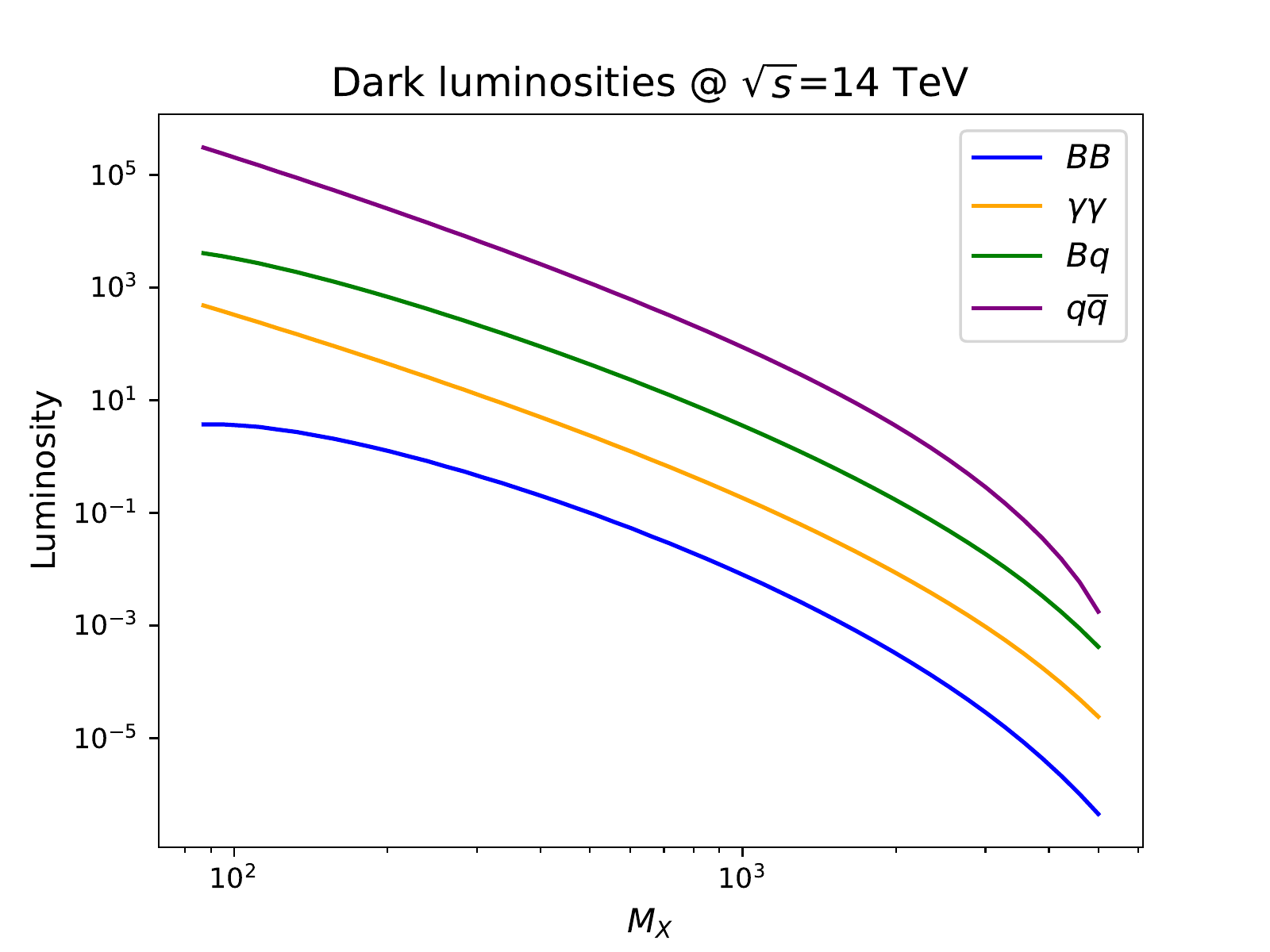}
\caption{Comparison of the absolute value of the $\Phi_{BB}$,
  $\Phi_{qB}$ central luminosities and the $\Phi_{\gamma\gamma}$  and
  $\Phi_{qq}$ central luminosities as a function of the invariant mass $M_X$
  at the centre of mass energy $\sqrt{s} = 14\ \text{TeV}$ for the dark PDF set obtained with 
 the dark photon coupling and mass set in Eq.~\eqref{eq:value}.
}
\label{fig:lowdarkmass}
\end{figure}
We first compare the size and the $M_X$-dependence of the different
parton luminosities in the candidate dark PDF set obtained by setting the mass and the coupling to the
values indicated in Eq.~\eqref{eq:value}. In Fig.~\ref{fig:lowdarkmass} we plot $\Phi_{BB},
\Phi_{Bq}$ as compared to $\Phi_{q\bar{q}}$,
$\Phi_{\gamma\gamma}$. We
observe that, while the $BB$ channel is suppressed by two powers of
the dark coupling, and its size never exceeds more than a fraction of a percent of the
$q\bar{q}$ luminosity, the $Bq$ channel grows from about 2\% of the
$q\bar{q}$ luminosity at $M_X\sim$ 1 TeV to about 8\% of the
$q\bar{q}$ luminosity at larger values of the invariant mass. Its
contribution exceeds that of $\gamma\gamma$ scattering by one order of magnitude. 

\begin{figure}[bht]
\centering
 \subfigure[$m_B=50\ \text{GeV}, \alpha_B = 3 \times 10^{-3}$]{
\includegraphics[scale=0.43]{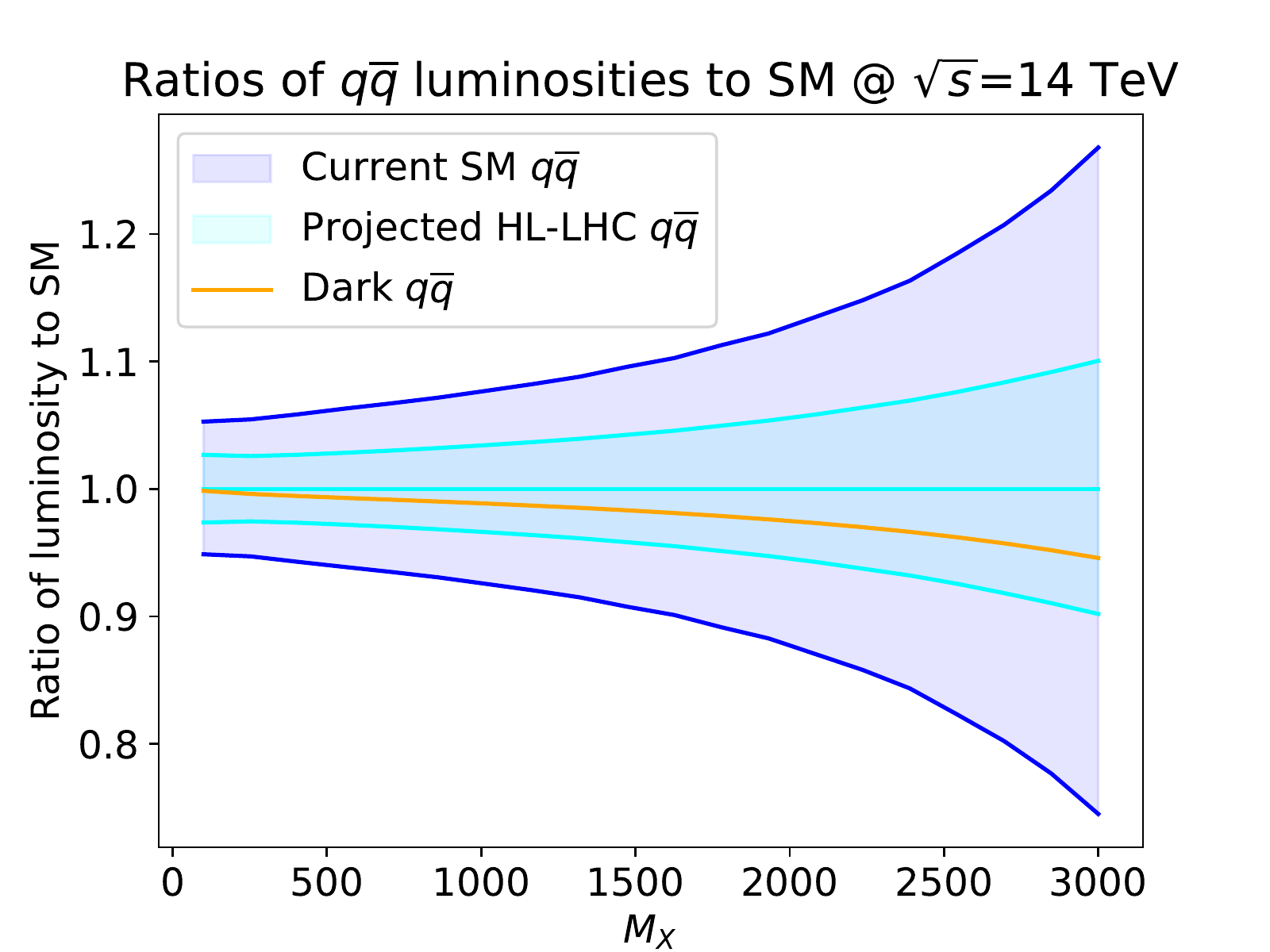}
}
\quad
 \subfigure[$m_B=50\ \text{GeV}, \alpha_B = 5 \times 10^{-3}$]{
\includegraphics[scale=0.43]{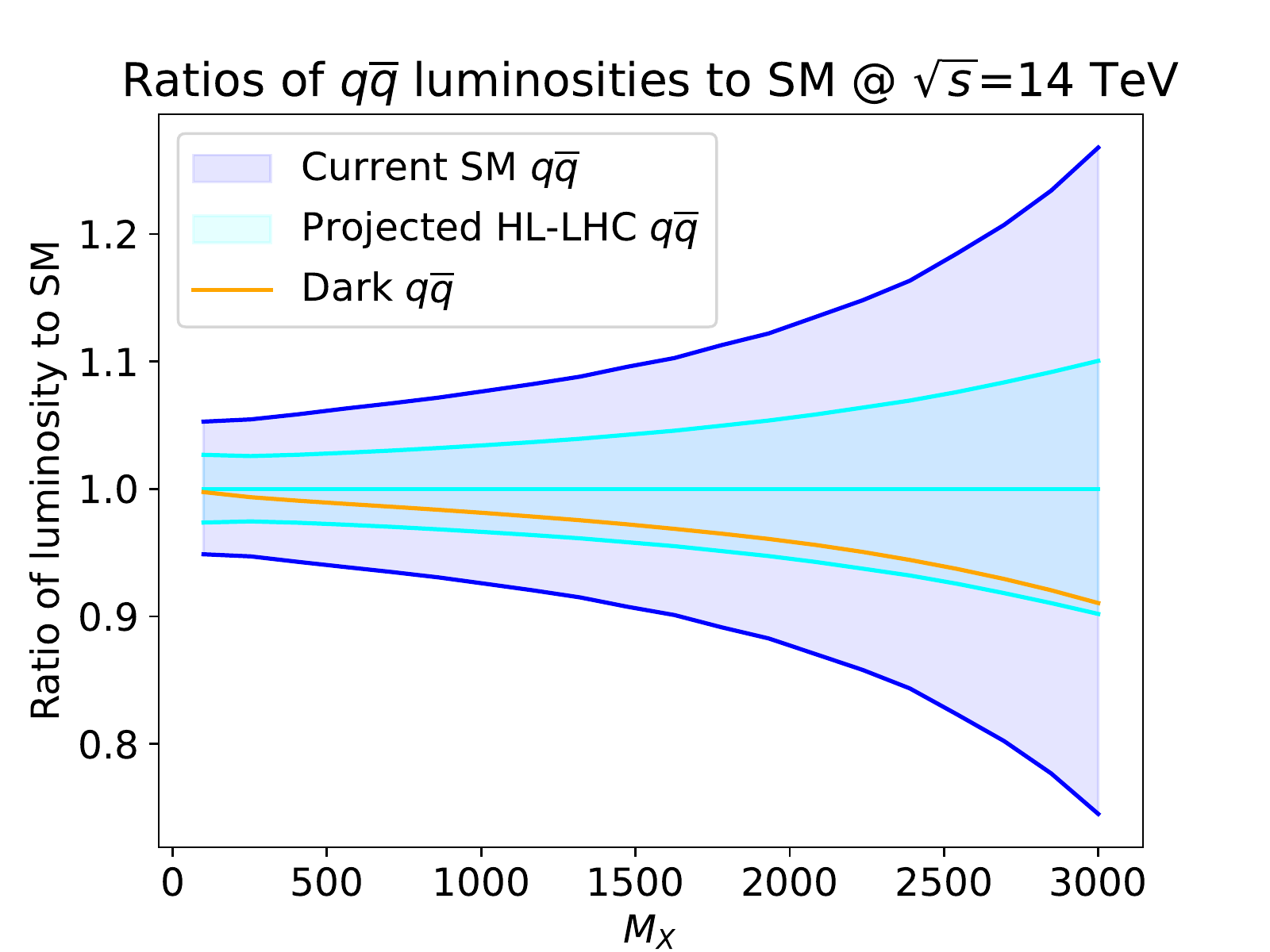}
}
\quad
  \subfigure[$m_B=5\ \text{GeV}, \alpha_B = 3 \times 10^{-3}$]{
 \includegraphics[scale=0.43]{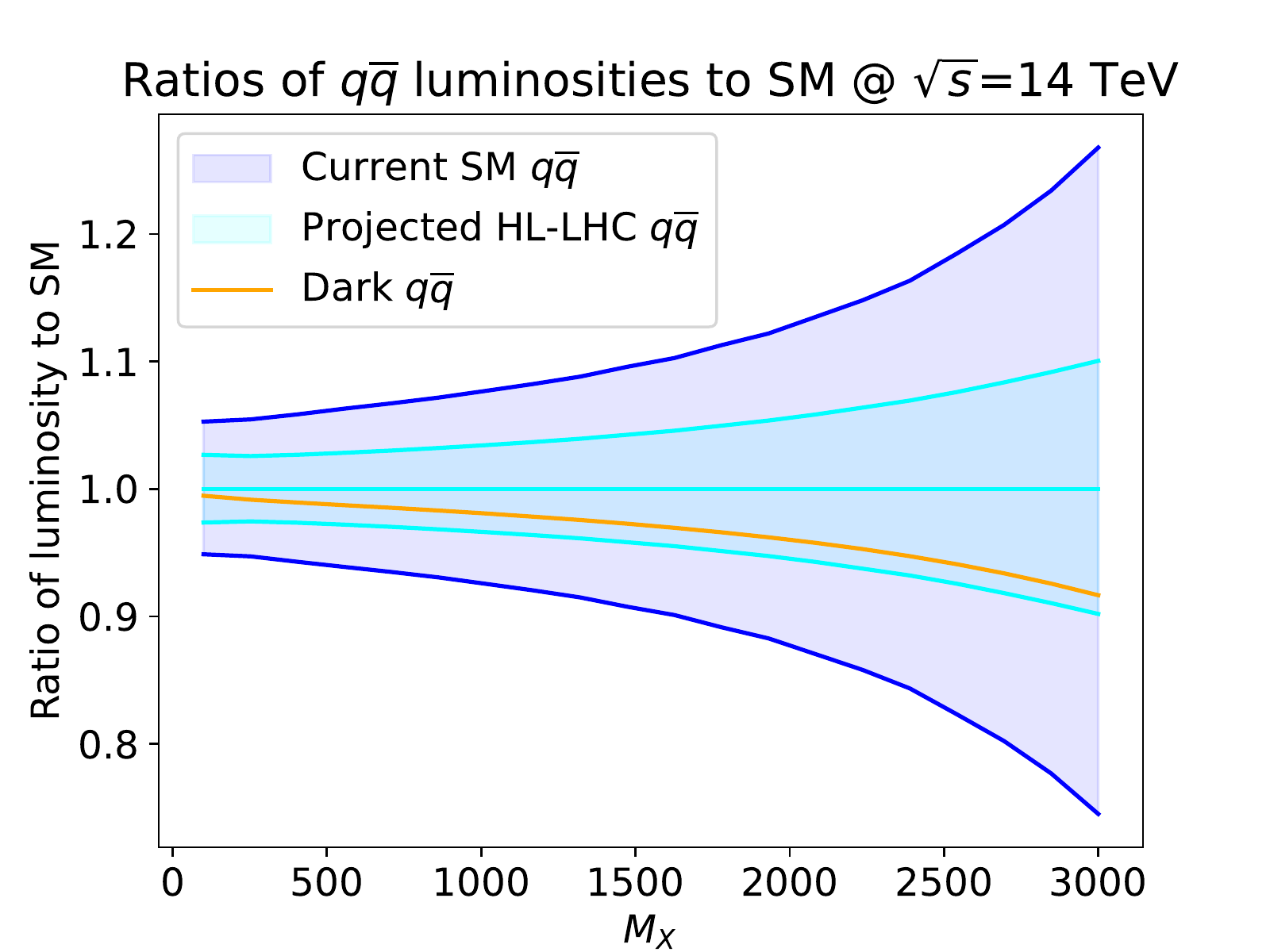}}
 \quad
  \subfigure[$m_B=5\ \text{GeV}, \alpha_B = 5 \times 10^{-3}$]{
 \includegraphics[scale=0.43]{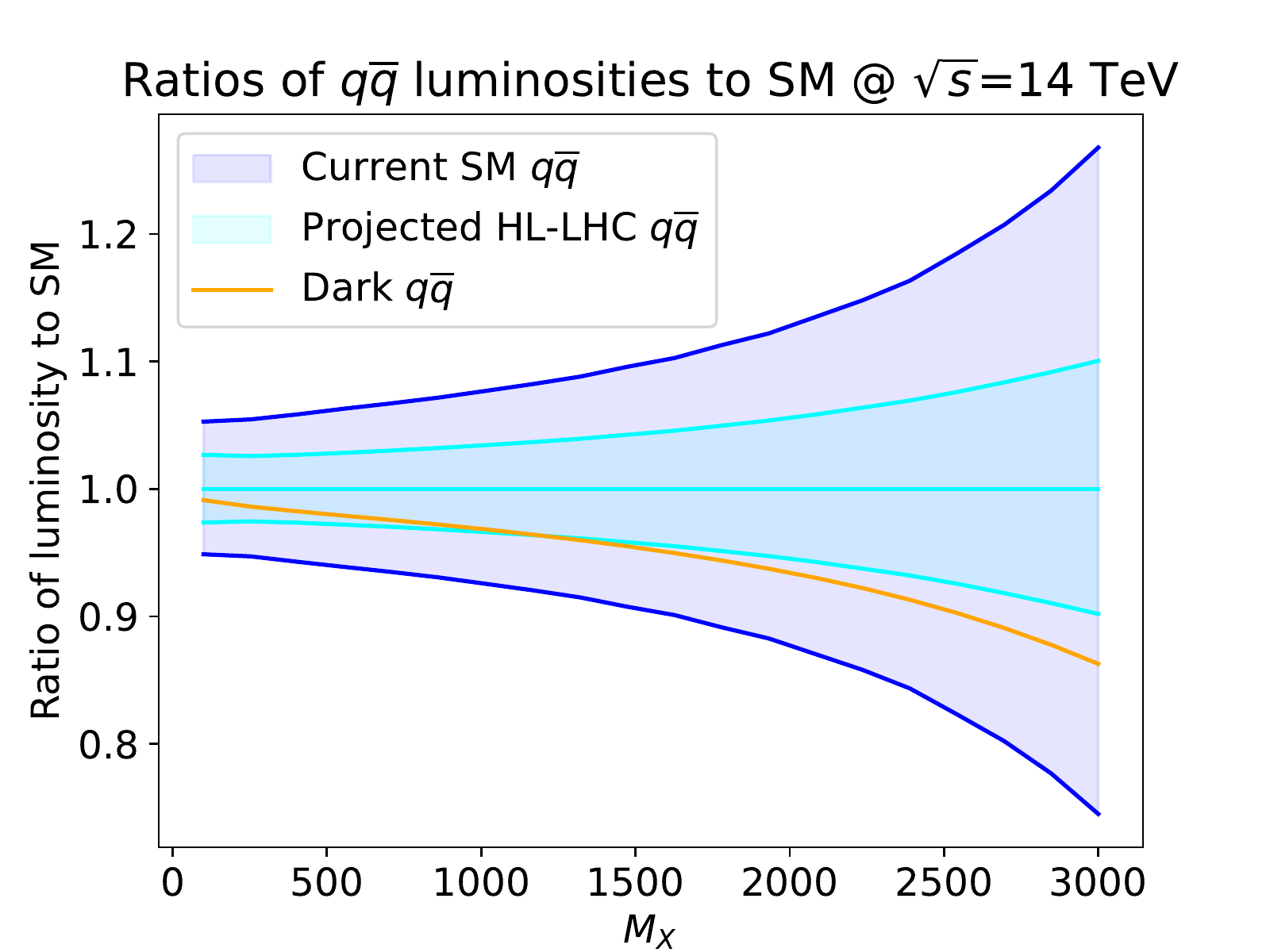}}
\caption{The ratio $\Phi_{q\bar{q}}^{\text{Dark}}/\Phi_{q\bar{q}}^{\text{SM}}$ for
  the total quark-anti-quark luminosity, at the centre of mass energy
  $\sqrt{S} = 14\ \text{TeV}$ for the values of mass and coupling
  indicated under each panel. 
  In each panel, the dark blue bands correspond to the current PDF
  uncertainty, while the light blue bands show the expected uncertainty on the
  PDF luminosity at the HL-LHC. See main text for more details.}
\label{fig:14tevlumis}
\end{figure}
We now turn to assess the change in the other luminosities, as a result of the
inclusion of a non-zero dark photon parton density.
In Fig.~\ref{fig:14tevlumis} we display the ratio of the dark-photon
modified quark-antiquark integrated luminosity
$\Phi_{q\bar{q}}^{\text{Dark}}$ with the baseline one, $\Phi_{q\bar{q}}^{\text{SM}}$ at the centre of mass energy
  $\sqrt{S} = 14\ \text{TeV}$, for different values of the $\alpha_B$
  and $m_B$ parameters, starting from our benchmark values, Eq.~\eqref{eq:value}.
  In each figure, the dark blue band corresponds the 68\% C.L. PDF
  uncertainty of the NNLO baseline {\tt NNPDF3.1luxQED} set, while the
  green bands show the projected PDF uncertainty on the
  parton luminosity at the HL-LHC; this estimate for the uncertainty
  on the PDF luminosity is obtained from the `optimistic' scenario, Scenario 3,
  analysed in \cite{Khalek:2018}, as above. 
  %
  Starting from the values of Eq.~\eqref{eq:value}, we observe that
  the deviation in the $q\bar{q}$ luminosity due to the presence of
  the dark photon is significant compared to the size of the
  projected PDF uncertainties at the HL-LHC. Decreasing the mass of
  the dark photon by a factor of 10 increases the impact of the dark
  photon on $q\bar{q}$ initiated observables, while increasing the
  coupling by less than a factor of 2 brings the luminosity beyond the
  edge of the 68\% C.L. error bands. 

Crucially, the effect of the dark photon is much larger in the $q\bar{q}$-initiated processes than in
any of the other channels, including $qq$, $qg$ and $gg$. This motivates the study of the high-mass Drell-Yan tails that we put
forward in the following section.

\subsection{Constraints from precise measurements
  of high-energy Drell-Yan tails}

Given that the $q\bar{q}$ channel is the most affected by the presence of a non-zero dark photon parton density, in this
study we focus on precise measurements of the high-mass Drell-Yan
tails at the HL-LHC. It is important to note that these projected data are not
included in the fit of the input PDF set used as a baseline,
otherwise, as was explicitly shown in ~\cite{Carrazza:2019sec,greljo_parton_2021,Iranipour:2022iak}, the interplay between the fit of the
new physics parameters and the fit of the PDF parametrisation at the
initial scale might distort the results. 

\begin{figure}
\centering
\includegraphics[width=0.9\textwidth]{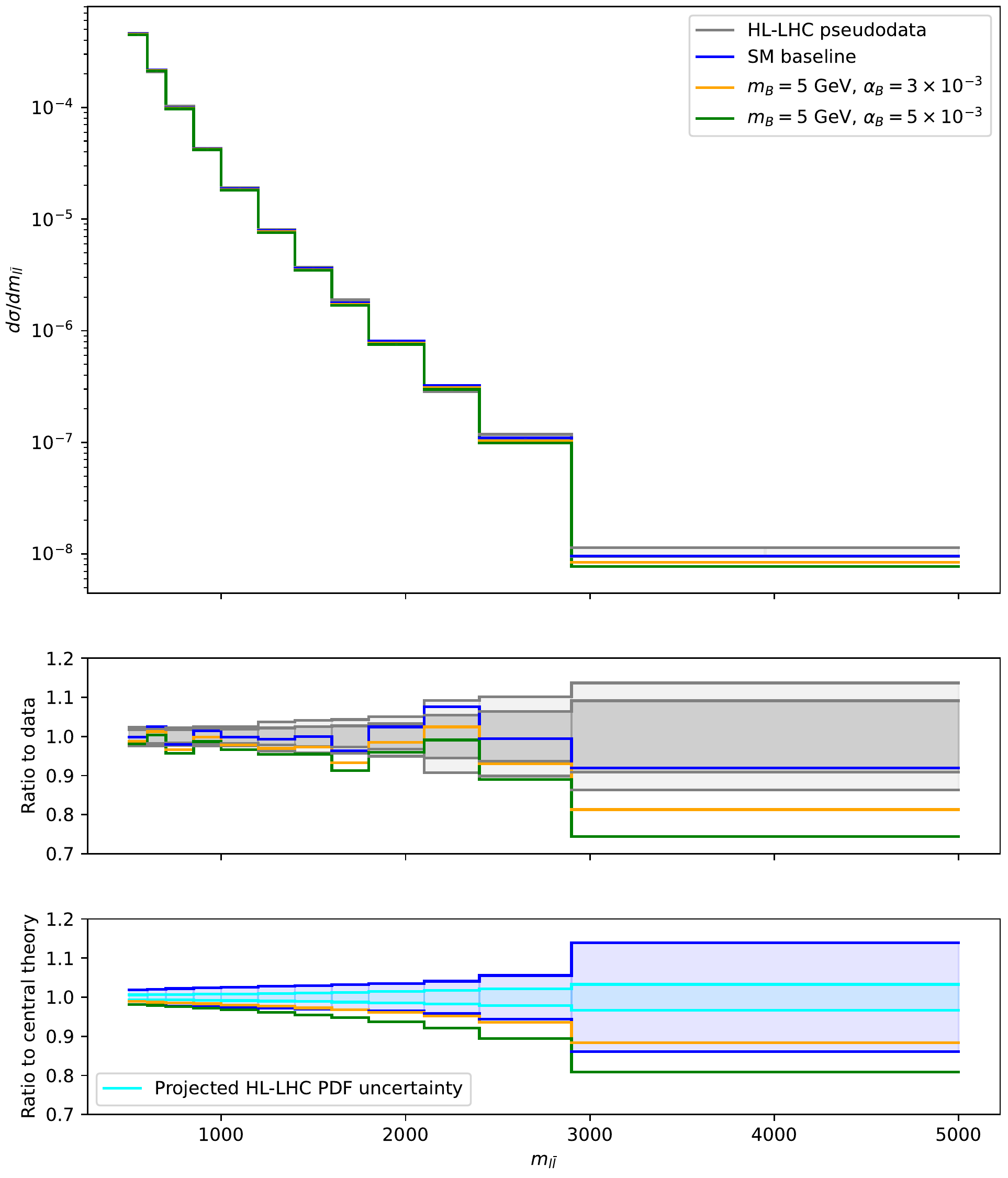}
\caption{\textbf{Top:} data-theory comparison between HL-LHC pseudodata in the electron channel generated according to
  Eq.~\eqref{eq:hllhc} (grey, with \textit{optimistic} uncertainties displayed), and the theoretical predictions obtained using
  the NNLO baseline PDF set {\tt NNPDF3.1luxQED} (blue) and
  those obtained using the dark PDF sets produced with parameters $(m_B,\alpha_B) = (5\ \text{GeV},3\times 10^{-3}), (5\ \text{GeV}, 5 \times 10^{-3})$ (yellow, green respectively).
\textbf{Middle:} ratio of the baseline SM central predictions obtained using the baseline, and the central predictions
  obtained using the two representative dark PDF sets, to the central values of the pseudodata.  The relative experimental
  uncertainties in both the {\it optimistic} scenario (dark grey) and
  {\it conservative} scenario (light grey) are displayed. \textbf{Bottom:} ratio of the central predictions obtained using the two
  representative dark PDF sets to the baseline SM central predictions, with both the PDF uncertainty from the baseline PDF set (dark blue) and the projected PDF uncertainty
  at the HL-LHC in the optimistic scenario of ~\cite{Khalek:2018}
  (light blue) displayed.
} 
\label{fig:data-theory}
\end{figure}

To generate the HL-LHC pseudodata for neutral-current
high-mass Drell-Yan cross sections at $\sqrt{S}=14$ TeV, we follow the procedure of~\cite{greljo_parton_2021}.
Namely, we adopt as reference the CMS  measurement at 13 TeV~\cite{CMS:2018mdl} based on $\mathcal{L}=2.8$ fb$^{-1}$.
The dilepton invariant mass distribution $m_{\ell\ell}$ is evaluated using the same selection
and acceptance cuts of~\cite{CMS:2018mdl}, but now with an extended
binning in $m_{\ell\ell}$ to account for the increase in luminosity.
We assume equal cuts for electrons and muons, and impose $|\eta_\ell|\le 2.4$,
$p_T^{\rm lead}\ge 20$ GeV, and $p_T^{\rm sublead}\ge 15$ GeV for the two
leading charged leptons of the event.
We restrict ourselves to  events with $m_{\ell\ell}$ greater than $500$ GeV, so that 
the total experimental uncertainty is not limited by our modelling
of the expected systematic errors, by making our projections unreliable.
To choose the binning, we require that the expected number of events per bin
is bigger than 30 to ensure the applicability of Gaussian statistics.
Taking into account these considerations, our choice of binning
for the $m_{\ell\ell}$ distributions at the HL-LHC both in the muon and electron channels 
are displayed in Fig.~\ref{fig:data-theory} with the highest energy bins reaching $m_{\ell\ell}\simeq 4 $ TeV.
In total, we have two invariant mass distributions of 12 bins each, one in the electron and one in the muon channels.

Concerning uncertainties, in ~\cite{greljo_parton_2021} this data is produced by assuming that the HL-LHC phase will
operate with a total integrated luminosity of $\mathcal{L} = 6\ \text{ab}^{-1}$
(from the combination of ATLAS and CMS, which provide ${\cal L}=3 \,{\rm ab}^{-1}$ each),
and also assuming a five-fold reduction in systematic uncertainty
compared to ~\cite{CMS:2018mdl}. 
We regard this scenario as {\it optimistic} in this paper; we also
manipulated the projected data so that it reflected a more {\it
  conservative} possibility, where the total integrated luminosity of
the high-mass Drell-Yan tail measurements is $\mathcal{L} = 3\
\text{ab}^{-1}$ (say, for example, they are made available only by
either ATLAS or CMS) and with a two-fold (rather than a five-fold) reduction in systematic uncertainties.

For these projections, the reference theory is the SM, with
theoretical predictions evaluated at NNLO in QCD including full 
NLO EW corrections (including in particular the photon-initiated
contributions); note, however, that the Drell-Yan production has been recently computed at
N${}^3$LO in QCD~\cite{Duhr:2020seh,Duhr:2021vwj}. In the kinematical
region that is explored by our HL-LHC projections ($m_{\ell\ell}>500$
GeV), the perturbative convergence of the series is good and
the N${}^3$LO computation is included within the NNLO prediction, with
missing higher order uncertainty going from about 1\% to a fraction of
a percent. Given the good perturbative convergence of
the matrix element calculation, and the absence of N${}^3$LO PDFs that match
the accuracy of the N${}^3$LO computation of the matrix element, we use the NNLO QCD and NLO EW accuracy of our calculations, both for the SM
baseline and for the dark-photon modified PDF set that we use to
compute the maximal sensitivity to the dark photon parameters. 

The central PDF set used as an input to generate the theoretical prediction is the SM baseline
that we use throughout the paper, namely the NNLO {\tt NNPDF3.1luxQED} set.
The central values for the HL-LHC pseudodata are then generated
by fluctuating the reference theory prediction by the expected total experimental
uncertainty, namely
\begin{equation}
  \label{eq:hllhc}
\sigma^{\rm hllhc}_{i} \equiv \sigma^{\rm th}_{i} \left( 1+ \lambda
  \delta_{\cal L}^{\rm exp} +  r_i\delta_{{\rm tot},i}^{\rm exp}   \right) \, , \qquad i=1,\ldots,n_{\rm bin} \, ,
\end{equation}
where $\lambda,r_i$ are univariate Gaussian random numbers, $\delta_{{\rm tot},i}^{\rm exp}$
is the total (relative) experimental uncertainty corresponding to this
specific bin
(excluding the luminosity and normalisation uncertainties), and $\delta_{\cal L}^{\rm exp}$
is the luminosity uncertainty, which is fully correlated amongst all
the pseudodata bins of the same experiment. We take this luminosity uncertainty to be
$\delta_{\cal L}^{\rm exp}=1.5$\%  for both ATLAS and CMS, as done in~\cite{Khalek:2018}.

To obtain bounds on the dark photon mass and coupling, we select a grid of benchmark
points $(m_B, \alpha_B)$ in the dark photon parameter space; our scan consists of $21$ points, distributed as a rectangular grid
with masses $m_B = 2, 5, 8, 10, 20, 50, 80\ \text{GeV}$ and couplings
$\alpha_B = 10^{-3}, 2\times10^{-3}, 3 \times 10^{-3}$. We then construct dark PDF sets at each 
of these benchmark points (thus a total of $21$ PDF sets, in each case including quarks, antiquarks, the gluon, the photon and the dark photon PDFs), using the 
appropriate values of $m_B, \alpha_B$, and hence
compute theoretical predictions in both the \textit{optimistic} and \textit{conservative} scenarios at each grid point. The predictions
are produced assuming that the primary contribution comes from the
$q\bar{q}$ channel; in particular, we note that the partonic diagrams that include a dark
photon in the initial state (such as $Bq\to \bar{q}l^+l^-$ or $B\bar{q}\to
ql^+l^-$) are suppressed by two powers of $\alpha_B$, one from the
dark photon PDF and one from the matrix element, and therefore are
suppressed beyond the accuracy of our calculation. 

In Fig.~\ref{fig:data-theory} we display the data-theory comparison
between the HL-LHC pseudodata in the electron channel, generated according to
Eq.~\eqref{eq:hllhc}, and both the SM theoretical predictions obtained using
  the NNLO baseline PDF set {\tt NNPDF3.1luxQED} and the predictions
  obtained using the dark PDF sets produced with the dark photon mass
  and coupling set
  to $(m_B=5\,{\rm GeV},\alpha_B=3\times 10^{-3})$ 
  and $(m_B=5\,{\rm GeV},\alpha_B=5\times 10^{-3})$ respectively. We also display 
 the ratio between the central values of those predictions and the
 central values of the pseudodata as compared to their relative
 experimental uncertainty in both the
 {\it optimistic} and {\it conservative} scenarios.
We see that whilst the SM predictions are within the $1\sigma$
experimental uncertainty (by construction), the dark-photon modified
predictions display significant deviations. 
In the bottom inset we show the ratio between the predictions
obtained in the two representative dark photon scenarios to the
central SM theoretical predictions obtained with the
the baseline SM PDF set. PDF uncertainties are shown; we display
both the current PDF uncertainty of the NNLO baseline PDF set {\tt
  NNPDF3.1luxQED} and the projected PDF uncertainties 
  at the HL-LHC, obtained as described at the beginning of this
  section. Comparing the size of the PDF uncertainties to the size of the
  projected experimental uncertainties at the HL-LHC, we observe
  that whilst the current PDF uncertainties are comparable to the
  experimental uncertainties of the projected data, the projected
  HL-LHC uncertainties are subdominant as compared to the
  experimental uncertainties of the pseudodata.

The $\chi^2$-statistic of the resulting dark PDF set's predictions on high-luminosity high-mass neutral-current DY data is defined as: 
\begin{equation}
\label{eq:chi2stat}
\chi^2(m_B, \alpha_B) := ||\vec{T}(m_B, \alpha_B) - \vec{D}||^2_{\Sigma^{-1}},
\end{equation}
where $||\vec{v}||_{A}^2 = \vec{v}^T A \vec{v}$, $\vec{D}$ is the projected data, 
$\vec{T}(m_B,\alpha_B)$ are the theoretical predictions using a dark
PDF set containing a dark photon of mass $m_B$ and coupling
$\alpha_B$, and $\Sigma$ is the total
covariance matrix (incorporating both experimental and theoretical
uncertainties):
\begin{equation}
\Sigma = \Sigma^{\rm th}+\Sigma^{\rm exp}.
\end{equation}
From Fig.~\ref{fig:data-theory} we observe that, depending on the assumption we make on PDF
uncertainties in the HL-LHC era, it may be important to
include the PDF uncertainties in the theory covariance matrix, while
the component of the theory covariance matrix associated with the
scale uncertainty of the NNLO computation is subdominant. 
Of course, it would be unrealistic to assume that the PDF uncertainty will not decrease as compared to the
  uncertainty of the {\tt NNPDF3.1luxQED} baseline, given that we already know
  that in the updated {\tt NNPDF4.0} set~\cite{Ball:2021leu} the uncertainty of the large-$x$
  quarks and antiquarks has already decreased by a sizeable amount
  thanks to the inclusion of precise LHC data. We thus decide to use
  the projected PDF uncertainties determined in~\cite{Khalek:2018}; in particular, we use Scenario 1 of ~\cite{Khalek:2018} (the conservative scenario) when we
  consider the {\it conservative} experimental scenario, and we use Scenario 3 of ~\cite{Khalek:2018} (the most optimistic scenario) when we
  consider the {\it optimistic} experimental scenario.  In Appendix C
  we discuss how our results depend on the assumptions we make on PDF
  uncertainties. 
  Assuming that the projected PDF uncertainties at the HL-LHC that we display in the
  bottom inset of Fig.~\ref{fig:data-theory} are realistic, even in
  the most optimistic scenario they still amount to 4\% to 6\% in the
  largest bins. Therefore, their contribution is much larger than the
  scale uncertainty of the Drell-Yan matrix element at NNLO in
  QCD; hence PDF uncertainty is the dominant theory uncertainty on the
   predictions, and thus it is this contribution that is included in the theory covariance matrix. 

To compute the contribution of PDF uncertainties to the theory
covariance matrix, we build the theoretical covariance as defined in~\cite{Hartland:2019bjb}:
\begin{equation}
\Sigma^{\rm th}_{ij} = \langle d\sigma_i^{{\rm th},(r)} d\sigma_j^{{\rm th},(r)}\rangle_{\rm rep} - \langle d\sigma_i^{{\rm th},(r)}\rangle_{\rm rep} \langle d\sigma_j^{{\rm th},(r)}\rangle_{\rm rep},
\end{equation}
where the theoretical predictions for the differential cross section $d\sigma_i^{{\rm th},(r)}$ are computed using the SM theory and the $r^{\rm th}$
replica from the baseline PDF set, with PDF uncertainties rescaled by
the HL-LHC uncertainty reduction, and averages $\langle \cdot \rangle_{\rm rep}$ are performed over the 
$N_{\rm rep} = 100$ replicas of this PDF set.

\begin{figure}
\centering
\includegraphics[width=0.95\textwidth]{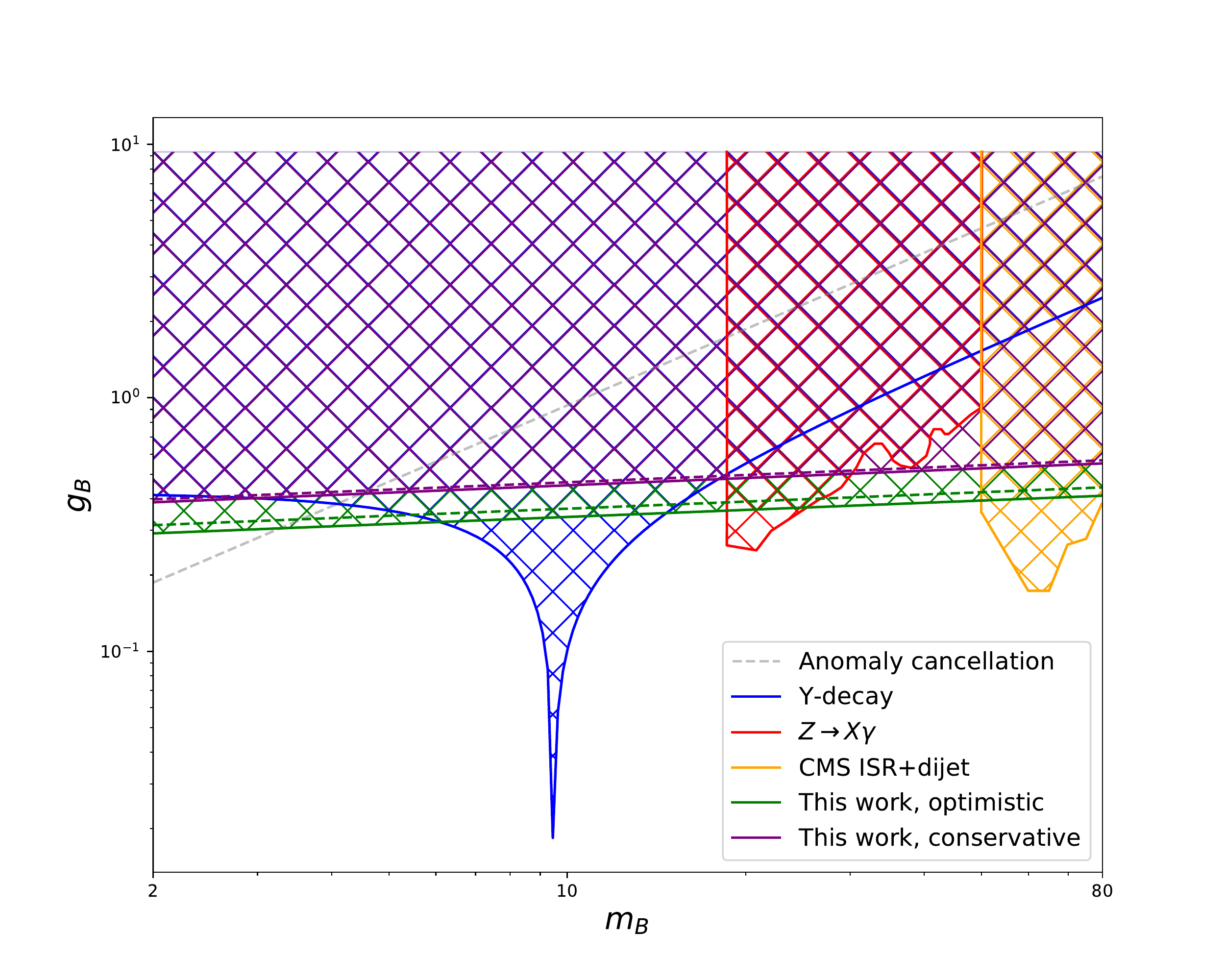}
\caption{Comparison of the projected HL-LHC sensitivity computed in
  this work in the optimistic (green) and conservative (purple) scenarios with the
  existing bounds described in Sec.~\ref{sec:existingconstraints}. The solid green and
  purple lines correspond to projected bounds obtained \textit{excluding} projected PDF uncertainty, whilst the 
  dashed lines correspond to projected bounds obtained \textit{including} projected PDF uncertainty, as discussed in
  the text.}
\label{fig:limits}
\end{figure}

We define the difference in $\chi^2$ to be:
\begin{equation}
\Delta \chi^2(m_B, \alpha_B) := \chi^2(m_B, \alpha_B) - \chi^2_0,
\end{equation}
where $\chi^2_0$ is the $\chi^2$-statistic when predictions from the baseline set are used instead. For each fixed $m_B = m_B^*$ in the scan, we then model $\Delta \chi^2(m_B^*,\alpha_B)$ as a quadratic in $\alpha_B$ and determine the point at which $\Delta \chi^2 = 3.8$, corresponding to a confidence of 95\% in a one-parameter scan. Hence, we construct 95\% confidence bounds on $m_B, \alpha_B$ (and hence $m_B, g_B$ via an appropriate conversion) as displayed in~Fig. \ref{fig:limits}. There, the purple (dashed) projected bounds are computed in the conservative scenario excluding (including) the PDF theory covariance matrix, and the green (dashed) projected bounds are computed in the optimistic scenario excluding (including) the PDF theory covariance matrix.

We observe that the projected HL-LHC sensitivity to the detection of a dark photon is competitive with existing experimental bounds, across a large range of possible masses, especially $m_B \in [2,6] \cup [25, 50]\ \text{GeV}$. Even in the most conservative scenario including PDF uncertainty (shown as a dashed purple line), the projected sensitivity remains competitive with experiment. Furthermore, one of the useful features of our projected sensitivity is that it is uniformly excluding across a large range (because of the logarithmic dependence on $m_B$); when compared to individual bounds one at a time, for example the $\Upsilon$-decay bounds, or the anomaly bounds, our projected sensitivity is powerful.
\section{Conclusions}
While the dark sector has evaded a non-gravitational detection for decades it is possible that the first clues may be right under our noses, hidden in the dark depths of the proton.  To illustrate this point we have considered the possible existence of a baryonic dark photon.  Being coupled to quarks it may, through radiative effects, be present in the proton and carry some fraction of its momentum at colliders.  In this work we have explicitly calculated the dark photon PDF and hence the dark parton luminosities at the LHC at leading order in the dark photon coupling.

Since any momentum fraction carried by the dark photon is not carried by the SM partons, the leading effect of the dark photon is to take away part of their momentum.  Importantly, this affects precision predictions for event distributions at colliders.  Being a high-precision observable at hadron colliders, DY production at high energies is sensitive to the dark photon content of the proton, through the reduction in light quark PDFs.  To this end, we have shown that future DY measurements at the HL-LHC are competitive with present lower energy probes in constraining regions of dark photon parameter space.  This reveals a new facet of the era of precision hadron collider physics we are currently entering.

Of course our projected sensitivity is based on the phenomenology tools that we currently have, while the actual analysis of real data at the HL-LHC will be able to use the tools that will be developed by then. Most importantly new global PDF sets will be made available, which will possibly be done at N$^3$LO, and resulting PDFs can be consistently used with N$^3$LO computations of the matrix elements. Moreover the associated PDF uncertainty will most certainly include a missing higher order uncertainty component in the PDF uncertainty that is not currently included in any of the global NNLO PDF fits. The actual sensitivity calculated with real data at the HL-LHC and the refined tools that the PDF community is working towards will be compared with the analysis we have put forward in this work, and the power of the LHC in constraining or revealing the presence of a dark photon will be unveiled. 

It is widely accepted that we must endeavour to search in every corner of parameter space for evidence of the dark sector.  This has led to a blossoming of novel theoretical ideas and ingenious experimental advances as an ever-greater range of viable dark sector possibilities come under scrutiny.  At the same it is important to reflect on our limited view of the dark sector and, perhaps, look inwards.  In this work we have pursued this strategy in a very literal sense, asking ``Could dark sector constituents lie within the baryons that make up the visible world?''  Our results suggest that they could, leaving their fingerprints in SM processes at the HL-LHC.

\section*{Acknowledgments}
We are very grateful to both Maeve Madigan and Cameron Voisey for supplying the 
`conservative' HL-LHC pseudodata, and the corresponding APPLgrids, used in this work.
We thank Valerio Bertone and Felix Heckhorn for their help with
APFEL. We thank Juan Rojo and Manuel Morales for their precious comments on the
manuscript. 

M.U. is supported by the European Research Council under the 
European Union's Horizon 2020 research and innovation Programme (grant agreement n.950246). 
M.U. is  partially supported by the STFC grant ST/T000694/1 and by the
Royal Society grant RGF/EA/180148.
The work of M.U. is also funded by the Royal Society grant DH150088. The work of J.M. is supported by the Sims Fund Studentship.

\newpage

\appendix
\section{Calculation of the dark photon splitting functions}
\label{app:splitting}
In this Appendix, we explicitly compute the
leading order dark photon splitting functions.

We begin by reminding the reader of one possible
definition of leading order splitting functions. Consider
a process initiated by some parton, which has the potential
to radiate another parton before participating in a
hard reaction (either a virtual parton, which is reabsorbed, or 
a real parton which is emitted). Suppose that radiation occurs via a 
vertex with coupling $g$ (which could be taken to be
$g_s$, $e$ or $g_B$, the strong, electric, or dark couplings
as appropriate), and corresponding fine
structure constant $\alpha = g^2/4\pi$.  The cross-section for
the radiative process can be shown to take the form:
\begin{equation}
\label{eq:splittingdefinition}
\sigma^{\text{NLO}} = \frac{\alpha}{2\pi} \int \frac{d(|\vec{l}_T|^2)}{|\vec{l}_T|^2} \int\limits_{0}^{1} dx\ P(x) \sigma^{\text{Born}}(xp) + \text{terms that are finite as $\vec{l}_T \rightarrow 0$}.
\end{equation}
\noindent Here, $\sigma^{\text{Born}}(xp)$ is the cross-section when parton radiation does not occur, $\vec{l}_T$ is the transverse momentum of the radiated parton and $x$ is the fraction of momentum of the parton that goes on to take part in the hard reaction. The factor $P(x)$ is called the \textit{splitting function} and depends on the type of parton initiating the process, the type of parton radiated, and the type of parton that goes on to participate in the hard reaction; typically it is written $P_{ij}(x)$ where $i$ is the parton which goes on to participate in the hard reaction, $j$ is the initial parton, and the third parton flavour is left implied by the structure of the theory. In particular, we see that it multiplies the most collinearly divergent part of the integrand on the right hand side of Eq.~\eqref{eq:splittingdefinition}.

We now explicitly compute the splitting functions for each of the four dark photon splitting channels, shown in Fig.~\ref{fig:splittingfunctions}.

\subsubsection*{Contribution to $P_{qq}^{(0,0,1)}(x)$}
\begin{figure}[H]
\centering
\begin{tikzpicture}
\begin{scope}[thick, decoration={
    markings,
    mark=at position 0.5 with {\arrow{>}}}
    ] 
\draw[postaction={decorate}] (-1,-1) node[below left] {$q$}-- (0,0);
\draw[dashed,postaction={decorate}] (-1,1) node[above left] {$k$}-- (0,0);
\end{scope}
\node[circle,fill=black,inner sep=0pt,minimum size=15pt] at (0,0) {};
\begin{scope}[ultra thick,decoration={
    markings,
    mark=at position 0.7 with {\arrow{>}}}
    ] 
\draw[postaction={decorate}] (0,0) -- (1,0) node[right] {$X$};
\end{scope}
\end{tikzpicture}
\caption{Feynman diagram corresponding to a Born-level quark-initiated hard process.}
\label{fig:borndiagram}
\end{figure}
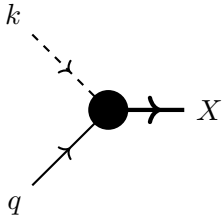

\noindent Consider a general leading order partonic process where a quark $q$ scatters off another particle $k$ to produce some final state $X$ via some hard interaction, as shown in Fig.~\ref{fig:borndiagram}.
\noindent The amplitude for this process can be written as $\mathcal{M}(p) u(p)$,
where $\mathcal{M}(p)$ is the amplitude for the unknown hard
scattering part of the diagram (taking as an argument the
four-momentum $p$ of our incoming quark $q$), including the final
state $X$ and the target particle $k$, and $u(p)$ is the appropriate plane-wave
spinor coming from the
initial quark $q$. After summing over all spins the amplitude gives rise to a cross-section:
\begin{equation}
\sigma^{\text{Born}}(p) = \frac{1}{4s} \sum_{X}  \mathcal{M}(p) \slashed{p} \mathcal{M}^{\dagger}(p),
\end{equation}
where the sum over $X$ includes an integral over the phase space of
the final state $X$.

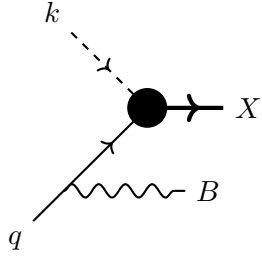
\begin{figure}[H]
\centering
\begin{tikzpicture}
\begin{scope}[thick,decoration={
    markings,
    mark=at position 0.5 with {\arrow{>}}}
    ] 
\draw[dashed,postaction={decorate}] (-1,1) node[above left] {$k$}-- (0,0);
\end{scope}
\draw[thick,snake it] (-1.1,-1.1) -- (0.5,-1.1) node[right]{$B$};
\begin{scope}[thick,decoration={
    markings,
    mark=at position 0.7 with {\arrow{>}}}
    ] 
\draw[postaction={decorate}] (-1.5,-1.5) node[below left] {$q$}-- (0,0);
\end{scope}
\node[circle,fill=black,inner sep=0pt,minimum size=15pt] at (0,0) {};
\begin{scope}[ultra thick,decoration={
    markings,
    mark=at position 0.7 with {\arrow{>}}}
    ] 
\draw[postaction={decorate}] (0,0) -- (1,0) node[right] {$X$};
\end{scope}
\end{tikzpicture}
\caption{Feynman diagram corresponding to real radiation of a dark photon.}
\label{fig:realradiation}
\end{figure}

On the other hand, the initial quark $q$ may also radiate a real dark photon before it participates in the interaction. Such a contribution arises from the diagram shown in Fig.~\ref{fig:realradiation}.
\noindent Let $l$ be the momentum of the outgoing dark photon. Then the amplitude for this diagram is:
\begin{equation}
\mathcal{M}(p-l) \frac{i}{\slashed{p} - \slashed{l}} \left(-\frac{ig_B}{3}\right) \slashed{\epsilon}(l) u(p) = \frac{g_B}{3(p-l)^2} \mathcal{M}(p-l) (\slashed{p} - \slashed{l}) \slashed{\epsilon}(l) u(p),
\label{eq:qqnloamplitude}
\end{equation}
\noindent where $\mathcal{M}(p-l)$ is the same amplitude for the hard
subgraph above, but this time with momentum $p-l$ entering the graph. $\epsilon(p)$ denotes the polarisation of the dark photon.

In order to compute the splitting function contribution $P_{qq}^{(0,0,1)}(x)$, we 
must determine the most collinearly-divergent part of the amplitude
of Eq.~\eqref{eq:qqnloamplitude}. To this end, we write $l$ in terms of its \textit{Sudakov decomposition}:
\begin{equation}
  \label{eq:sudakov}
l = (1-x)p + l_T + \frac{|\vec{l}_T|^2}{1-x} \eta,
\end{equation}
where $l_T$ is some transverse four-vector, $\eta$ is a null vector satisfying $2 \eta \cdot p = 1$, and
$x$ is a constant chosen to make the decomposition true. In particular, we see that as $l_T \rightarrow 0$, we have $l \rightarrow (1-x)p$, i.e. the initial and radiated partons become collinear, and the parton which goes on to participate in the hard interaction has momentum fraction $x$. Therefore, in this notation, the collinear limit is $l_T \rightarrow 0$.

We now write Eq.~\eqref{eq:qqnloamplitude} in terms of $l_T$, and hence determine the most collinearly divergent parts that we are required to retain. Via a short calculation,
we can obtain:
\begin{equation}
-\frac{g_B (1-x)}{3 |\vec{l}_T|^2} \mathcal{M}(p-l) \left(x \slashed{p} - \slashed{l}_T - \frac{|\vec{l}_T|^2}{1-x}\slashed{\eta} \right) \slashed{\epsilon}(l) u(p).
\end{equation}
Retaining only the divergent terms as $l_T \rightarrow 0$, the modulus-squared spin-sum/averaged amplitude is then given by:
\begin{equation}
  \frac{g_B^2}{9 |\vec{l}_T|^2} \left( \frac{1 + x^2}{x} \right) \mathcal{M}(xp) x\slashed{p} \mathcal{M}^{\dagger}(xp).
\end{equation}
Inserting into the standard cross-section formula, and appropriately changing integration variables, reveals that the most collinearly-divergent part of the cross-section
associated to this process is given by:
\begin{equation}
\sigma^{\text{NLO,real}}(p) = \frac{\alpha_B}{2 \pi} \int \frac{d(|\vec{l}_T|^2)}{|\vec{l}_T|^2} \int\limits_{0}^{1} dx \ \frac{1}{9} \left( \frac{1+x^2}{1-x} \right) \sigma^{\text{Born}}(xp),
\end{equation}
As it stands, we see that $\sigma^{\text{NLO,real}}(p)$ also contains a \textit{soft divergence} as $x \rightarrow 1$,
which characterises the divergence that occurs when the radiated
parton's energy tends to zero. The soft divergent part is cancelled by adding up the virtual
contribution, which corresponds to a dark photon being emitted by the quark, and later
being reabsorbed. 

Fortunately, there is a clever probability argument which sidesteps
actual computation of the virtual corrections. Let's begin by writing the virtual corrections as:
\begin{equation}
\sigma^{\text{NLO,virtual}}(p) = A \sigma^{\text{Born}}(p),
\end{equation}
where $A$ is some appropriately chosen constant. It follows that the sum of the virtual corrections and the leading order graph is given by $(1+A)\sigma^{\text{Born}}(p)$. Thus adding the virtual corrections, leading order graph and the real emission graph, we have:
\begin{equation}
\sigma^{\text{total}}(p) = \int\limits_{0}^{1} dx \left( \frac{1}{9} \left( \frac{1+x^2}{1-x} \right) + (1+A) \delta(1-x) \right) \sigma^{\text{Born}}(xp).
\end{equation}
We interpret the coefficient of $\sigma^{\text{Born}}(xp)$ as the possible outcomes for the initial quark: it either radiates a $B$-boson, radiates a virtual quark which later recombines with the original quark, or it does none of the above. At this order, these are the \textit{only} possibilities, and hence the probabilities of these events must sum to $1$:
\begin{equation}
\int\limits_{0}^{1} dx \left( \frac{1}{9}\left( \frac{1 + x^2}{1-x}\right) \Theta(1 - \epsilon - x) + (1 + A)\delta(1-x) \right) = 1.
\end{equation}
where we introduce the regulator $\epsilon$ that cuts off the integral
at $1-\epsilon$ by using the Heaviside step function $\Theta$. We
can now compute the (divergent) value of $A$ that enforces this
condition, finding:
\begin{equation}
A = \frac{1}{6} + \frac{2}{9} \log(\epsilon).
\end{equation}
This reveals that the splitting function can be written (in a distributional sense) as:
\begin{equation}
P_{qq}^{(0,0,1)}(x) = \lim_{\epsilon \rightarrow 0} \left[\frac{1}{9}\left( \frac{1 + x^2}{1-x}\right) \Theta(1 - \epsilon - x) + \left( \frac{1}{6} + \frac{2}{9} \log(\epsilon)\right)  \delta(1-x)\right].
\end{equation}
Manipulating this distribution shows that $P_{qq}^{(0,0,1)}(x)$ can be
written in the more concise form:
\begin{equation}
P_{qq}^{(0,0,1)}(x) = \frac{1 + x^2}{9(1-x)_+} + \frac{1}{6} \delta(1-x),
\end{equation}
where the \textit{plus distribution} is defined in Eq.~\eqref{eq:plusdistribution}.

\subsubsection*{Contribution to $P_{qB}^{(0,0,1)}(x)$}
\noindent In order to work out the contribution to $P_{qB}^{(0,0,1)}(x)$, we consider the radiation of an antiquark from a dark photon, as shown in Fig.~\ref{fig:antiquarkradiation}.
\begin{figure}[H]
\centering
\begin{tikzpicture}
\begin{scope}[thick,decoration={
    markings,
    mark=at position 0.5 with {\arrow{>}}}
    ] 
\draw[dashed,postaction={decorate}] (-1,1) node[above left] {$k$}-- (0,0);
\draw[postaction={decorate}] (0.5,-0.8) node[right] {$\bar{q}$} -- (-0.8,-0.8);
\end{scope}
\draw[thick,snake it] (-1.5,-1.5) node[below left]{$B$}-- (-0.8,-0.8) ;
\begin{scope}[thick,decoration={
    markings,
    mark=at position 0.5 with {\arrow{>}}}
    ] 
\draw[postaction={decorate}] (-0.8,-0.8)-- (0,0);
\end{scope}
\node[circle,fill=black,inner sep=0pt,minimum size=15pt] at (0,0) {};
\begin{scope}[ultra thick,decoration={
    markings,
    mark=at position 0.7 with {\arrow{>}}}
    ] 
\draw[postaction={decorate}] (0,0) -- (1,0) node[right] {$X$};
\end{scope}
\end{tikzpicture}
\caption{Feynman diagram corresponding to real radiation of an antiquark.}
\label{fig:antiquarkradiation}
\end{figure}
\noindent The amplitude for this diagram is given by:
\begin{equation}
\frac{g_B}{3(p-l)^2} \mathcal{M}(p-l) (\slashed{p} - \slashed{l}) \slashed{\epsilon}(p) v(l).
\end{equation}
As before, we can use the Sudakov decomposition to write the amplitude in terms of the transverse momentum $l_T$:
\begin{equation}
-\frac{g_B (1-x)}{3 | \vec{l}_T|^2} \mathcal{M}(xp) (x \slashed{p} - \slashed{l}_T) \slashed{\epsilon}(p) v(l),
\end{equation}
where we drop all terms that remain finite as $l_T \rightarrow 0$.
Further simplification of the spinor structure, and ignoring terms that remain finite as $l_T \rightarrow 0$ throughout, 
the modulus-squared spin-sum/averaged amplitude becomes:
\begin{equation}
\frac{g_B^2(1-x)}{9 |\vec{l}_T|^2} \left(\frac{2x^2 - 2x + 1}{x}\right)\mathcal{M}(xp) x \slashed{p} \mathcal{M}^{\dagger}(xp).
\end{equation}
Inserting into the cross-section formula, we have:
\begin{equation}
\sigma^{\text{NLO}}(p) = \frac{\alpha_B}{2\pi} \int \frac{d(|\vec{l}_T|^2)}{|\vec{l}_T|^2} \int\limits_{0}^{1} dx \ \left(\frac{2x^2 - 2x + 1}{9} \right) \sigma^{\text{Born}}(xp).
\end{equation}
and hence we deduce that the splitting function is given by:
\begin{equation}
P_{qB}^{(0,0,1)}(x) = \frac{2x^2 -2x + 1}{9} = \frac{(1-x)^2 + x^2}{9}.
\end{equation}
In this case, there are no virtual corrections to take care of at this order.

\newpage
\subsubsection*{Contribution to $P_{BB}^{(0,0,1)}(x)$}
\noindent Having found $P_{qB}^{(0,0,1)}(x)$, it is straightforward to find $P_{BB}^{(0,0,1)}(x)$ using the fact that probabilities must sum to $1$. At this order, the only contribution to $P_{BB}^{(0,0,1)}(x)$ is given by a virtual correction to the incoming dark photon, namely a quark loop on this line. Therefore, noting that the probability of an incoming dark photon splitting plus the probability of an incoming dark photon staying intact must sum to $1$, we have:
\begin{equation}
\int\limits_{0}^{1} dx\ \left( \frac{x^2 + (1-x)^2}{9} + (1 + A)\delta(1-x)\right) = 1,
\end{equation}
\noindent where $A$ denotes the virtual contribution and the $1$ denotes the contribution from the Born amplitude for the dark photon initiated process. Solving this equation, we find that $A = -2/27$, and it follows that the splitting function is given by:
\begin{equation}
P_{BB}^{(0,0,1)}(x) = -\frac{2}{27} \delta(1-x).
\end{equation}

\subsubsection*{Contribution to $P_{Bq}^{(0,0,1)}(x)$}
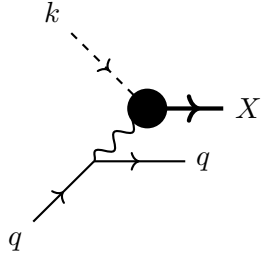
\begin{figure}[H]
\centering
\begin{tikzpicture}
\begin{scope}[thick,decoration={
    markings,
    mark=at position 0.5 with {\arrow{>}}}
    ] 
\draw[dashed,postaction={decorate}] (-1,1) node[above left] {$k$}-- (0,0);
\end{scope}
\draw[thick,snake it] (-0.7,-0.7) -- (0,0);
\begin{scope}[thick,decoration={
    markings,
    mark=at position 0.5 with {\arrow{>}}}
    ] 
\draw[postaction={decorate}] (-1.5,-1.5) node[below left] {$q$}-- (-0.7,-0.7);
\draw[postaction={decorate}] (-0.7,-0.7) -- (0.5,-0.7) node[right]{$q$};
\end{scope}
\node[circle,fill=black,inner sep=0pt,minimum size=15pt] at (0,0) {};
\begin{scope}[ultra thick,decoration={
    markings,
    mark=at position 0.7 with {\arrow{>}}}
    ] 
\draw[postaction={decorate}] (0,0) -- (1,0) node[right] {$X$};
\end{scope}
\end{tikzpicture}
\caption{Feynman diagram corresponding to real radiation of a quark.}
\label{fig:quarksplitting}
\end{figure}

\noindent Finally, to obtain $P_{Bq}^{(0,0,1)}(x)$, we consider the radiation of a dark photon from a quark, as displayed in Fig.~\ref{fig:quarksplitting}.
\noindent Again, the splitting function can be found using a simple probability argument. We note that:
\begin{equation}
P_{Bq}^{(0,0,1)}(x) = P_{qq}^{(0,0,1)}(1-x) - \text{virtual corrections}.
\label{eq:probargument}
\end{equation}
\noindent To see why this equation holds, recall that $P_{qq}^{(0,0,1)}(x)$ is the probability that a quark of momentum $p$ will split into a quark of momentum $xp$ (which participates in a hard interaction) and a dark photon of momentum $(1-x)p$, whilst $P_{Bq}^{(0,0,1)}(1-x)$ is the probability that a quark of momentum $p$ will split into a dark photon of momentum $xp$ (which participates in a hard interaction) and a dark photon of momentum $(1-x)p$. The only difference is the participant in the hard reaction, whose only effect on the splitting function is to determine whether virtual corrections are necessary or not. Hence Eq.~\eqref{eq:probargument} follows.

Therefore, discarding the virtual corrections from $P_{qq}^{(0,0,1)}(x)$ which we computed earlier, we immediately have:
\begin{equation}
P_{Bq}^{(0,0,1)}(x) = \frac{1}{9} \left( \frac{1 + (1-x)^2}{1 - (1-x)} \right) = \frac{1}{9} \left( \frac{1 + (1-x)^2}{x} \right).
\end{equation}

\section{Code implementation of the coupled evolution} 
\label{app:code}
The solution of the DGLAP equations in the presence of QED corrections~\cite{Spiesberger:1994dm,Roth:2004ti,Martin:2004dh} has been implemented in several tools; 
in particular, they are implemented in \texttt{APFEL}~\cite{Bertone:2013vaa}, which provides a public 
code, accurate and flexible, that can be used to perform PDF evolution up to NNLO
in QCD and NLO in QED, using a variety of heavy flavours schemes.

We implemented the modified DGLAP equations~\eqref{eq:basicdglap} in {\tt APFEL}. 
The evolution is performed in $x$-space (rather than Mellin $N$-space), and uses a rotated 
basis of PDFs such that a maximal number of PDF flavour combinations
evolve independently. If we define the following vector of PDFs:
\begin{equation}
\label{eq:singlet}
\vec{q}^S = \begin{pmatrix} g \\ \gamma \\ \Sigma \\ \Delta_{\Sigma} \\ B\end{pmatrix},
\end{equation}
where:
\begin{equation}
\label{eq:singlet2}
\Sigma = \sum_{f=u,d,s,c} (f + \bar{f}), \qquad \Delta_{\Sigma} = \sum_{f=u,c} (f + \bar{f}) - \sum_{f=d,s} (f + \bar{f}),
\end{equation}
then we can choose further independent flavour combinations of PDFs, spanning the
complete space of PDFs, such that all of the remaining flavour combinations' evolution equations decouple;
 this greatly simplifies the computational work. The remaining matrix equation for $\vec{q}^S$ can 
 be shown to take the form:
 \renewcommand{\arraystretch}{0.8}
\begin{equation}
\label{eq:dglapevolution}
Q^2 \frac{\partial \vec{q}^S}{\partial Q^2} = \left(\begin{array}{cccc|c} & & & &  0 \\ & \ddots& \ddots& & 0 \\ & \ddots& \ddots& & P_{q B} \\[1ex] & & & & P_{q B} \\[1ex] \hline 0 & 0 & P_{Bq} & 0 & P_{BB} \end{array}\right) \otimes \vec{q}^S.
\end{equation}
Here, the dots denote the relevant SM matrix, with the quark-quark splitting function corrected with a dark contribution as appropriate. This equation (together with the other decoupled scalar equations) is solved using an adaptive step-size fifth-order Runge-Kutta method, 
as described in~\cite{Bertone:2013vaa}.

To solve the modified DGLAP equations~\eqref{eq:basicdglap}, we must
also specify initial conditions for the dark photon;
this is where we make appropriate ans\"{a}tze for the functional form of the dark photon at the initial scale $Q_0 = 1.65\ \text{GeV}$. 
If the mass of the dark photon $m_B$ were less than the scale $Q_0$, we
could postulate a functional form for the initial dark photon PDF
assuming that the dark photon PDF is primarily generated by quark
splitting.
An appropriate initial condition in this case would be given by:
\begin{equation}
\label{eq:lowmassansatz}
B(x,Q_0^2) = \frac{\alpha_B}{2\pi} \log\left( \frac{Q_0^2}{m_B^2} \right) \sum_{j=1}^{n_f} (P_{Bq_j} \otimes (q_j(x,Q_0^2)+ \bar{q}_j(x,Q_0^2)),.
\end{equation}
On the other hand, in our phenomenologically relevant region, we have $m_B >
2$ GeV; thus in our case, we always have $m_B > Q_0$. As a result, we set $B(x,Q^2) = 0$ for all $Q < m_B$ and we generate the dark photon PDF dynamically at the threshold $Q = m_B$ from PDF evolution, similar to the treatment of heavy quarks~\cite{Maltoni:2012pa,Bertone:2017djs}, and the tauon PDF in~\cite{Bertone:2015lqa}.


\section{PDF uncertainties}

\begin{table}[tb]
\centering
\begin{tabularx}{\textwidth}{X|C{1.8cm}C{1.8cm}C{1.8cm}C{1.8cm}}
\toprule
 Assumption &  A &   B & C & D \\
\midrule
$m_B$ &\multicolumn{4}{c}{Optimistic scenario } \\
\midrule
5 GeV   & 0.317  & 0.341  & 0.356 & 0.597 \\
50 GeV & 0.393  & 0.424 & 0.443  & 0.758 \\
\midrule
$m_B$ &\multicolumn{4}{c}{Conservative scenario } \\
\midrule
5 GeV   & 0.421   & 0.428 & 0.433  & 0.622 \\
50 GeV &  0.526  & 0.535 & 0.542  & 0.791 \\
\bottomrule
\end{tabularx}

\caption{A comparison between the shifts in the 95\% C.L.~sensitivity
obtained by making different assumptions on PDF uncertainty reduction
in the HL-LHC phase. See the text for the definition of the scenarios.}
\label{tab:PDFunc}
\end{table}

Here we quantify the effect of PDF uncertainty. PDF
uncertainties have an experimental component and a theoretical one; at
the moment, the highest perturbative order that is included by default
in a PDF fit is NNLO (given that the PDF evolution at N${}^3$LO is not
fully known yet), and as of yet no PDF set includes the component of
the PDF uncertainty associated with missing higher order uncertainties
in the theory predictions used in the fit. We have
indications~\cite{NNPDF:2019ubu} that the inclusion of theory
uncertainties in the PDF fits will not significantly enhance the PDF
uncertainties in phenomenologically relevant observables. Hence, in this
study, we make assumptions on the PDF uncertainties based on
the best tools that we have available. There are several assumptions
that can be made:
\begin{itemize}
\item \textbf{Assumption A:} Ignore PDF uncertainties.
\item \textbf{Assumption B:} Assume that the correct estimate for PDF
  uncertainties at the end of the HL-LHC phase is the one given in
  Scenario 3 of Ref.~\cite{Khalek:2018}.
  \item \textbf{Assumption C:} Assume that the correct estimate for PDF
    uncertainties at the end of the HL-LHC phase is the one given in
    Scenario 1 of Ref.~\cite{Khalek:2018}. This is the least
    optimistic projections for PDF uncertainties.
    \item \textbf{Assumption D:} Assume that the PDF uncertainties will not
      decrease in the next 15-20 years from the current level, hence
      use the PDF uncertainty associated with the {\tt NNPDF3.1luxQED}
      NNLO baseline. 
    \end{itemize}
  In this paper, we adopt `Assumption B' for
  our {\it optimistic} scenario, and we adopt `Assumption C' for
  our {\it conservative} scenario. In Table~\ref{tab:PDFunc}, we report
  the maximal 95\% sensitivity obtained by varying these
  assumptions from our default choice.
We observe that the uncertainties associated
with `Assumption D' would spoil the sensitivity, but this scenario is 
extremely far from realistic.


\providecommand{\href}[2]{#2}\begingroup\raggedright\endgroup

\end{document}